\documentclass[twocolumn]{aastex63}

\newcommand{\ctext}[1]{\raise0.2ex\hbox{\textcircled{\scriptsize{#1}}}}

\newcommand{\todo}{\ifmmode \text{\color{purple}\Huge{\(\bullet\)}} \else {\color{purple}{\Huge$\bullet$}}\fi}
\newcommand{\finish}{\ifmmode \text{\color{blue}\Huge{\(\bullet\)}} \else {\color{blue}{\Huge$\bullet$}}\fi}

\usepackage{makecell}
\usepackage{lineno}

\accepted{May 27, 2026}
\submitjournal{ApJS}

\usepackage{tabularx} 
\usepackage{xurl} 
\usepackage{float} 

\shorttitle{WERGS}
\shortauthors{Uchiyama et al.}

\begin{document}

\title{A Wide and Deep Exploration of Radio-detected Active Galactic Nuclei with Subaru HSC (WERGS). XII.  Final Optical Identification of VLASS Radio Sources from the Subaru/HSC-SSP Wide Survey Over 1200~deg$^2$}
\correspondingauthor{Hisakazu Uchiyama, Kohei Ichikawa}
\email{hisakazu.uchiyama.86@hosei.ac.jp, k.ichikawa@astr.tohoku.ac.jp}

\author[0000-0002-0673-0632]{Hisakazu Uchiyama}
\affiliation{Department of Advanced Sciences, Faculty of Science and Engineering, Hosei University, Koganei, Tokyo 184-8584, Japan} 
\affiliation{National Astronomical Observatory of Japan, Mitaka, Tokyo 181-8588, Japan} 

\author[0000-0002-4377-903X]{Kohei Ichikawa}
\affiliation{Frontier Research Institute for Interdisciplinary Sciences, Tohoku University, Sendai, Miyagi 980-8578, Japan}
\affiliation{Astronomical Institute, Tohoku University, Aramaki, Aoba, Sendai 980-8578, Japan} 

\author[0009-0002-4201-7727]{Youwen Kong}
\affiliation{Institute of Astronomy, Graduate School of Science, The University of Tokyo, 2-21-1 Osawa, Mitaka, Tokyo, 181-0015 Japan}

\author[0009-0001-3910-2288]{Yuxing Zhong}
\affiliation{Department of Physics, School of Advanced Science and Engineering, Faculty of Science and Engineering, Waseda University, 3-4-1, Okubo, Shinjuku, Tokyo 169-8555, Japan}

\author[0000-0003-2682-473X]{Xiaoyang Chen}
\affiliation{Frontier Research Institute for Interdisciplinary Sciences, Tohoku University, Sendai, Miyagi 980-8578, Japan}

\author{Tohru Nagao}
\affiliation{Research Center for Space and Cosmic Evolution, Ehime University, Bunkyo-cho 2-5, Matsuyama, Ehime 790-8577, Japan}

\author[0000-0003-4814-0101]{Kianhong Lee}
\affiliation{Department of Physics, Graduate School of Science, Nagoya University, Furo, Chikusa, Nagoya, Aichi 464-8602, Japan}

\author{Kotaro Kohno}
\affiliation{Institute of Astronomy, Graduate School of Science, The University of Tokyo, Mitaka, Tokyo 181-0015, Japan}
\affiliation{Research Center for the Early Universe, Graduate School of Science, The University of Tokyo, 7-3-1 Hongo, Bunkyo-ku, Tokyo 113-0033, Japan}
\affiliation{ILANCE, CNRS – University of Tokyo International Research Laboratory, Kashiwa, Chiba 277-8582, Japan}

\author[0000-0003-2213-7983]{Bovornpratch Vijarnwannaluk}
\affiliation{Academia Sinica Institute of Astronomy and Astrophysics, 11F of Astronomy-Mathematics Building, AS/NTU, No.1, Section 4, Roosevelt Road, Taipei 10617, Taiwan}
\affiliation{Astronomical Institute, Tohoku University, Aramaki, Aoba, Sendai 980-8578, Japan}

\author{Masayuki Akiyama}
\affiliation{Astronomical Institute, Tohoku University, Aramaki, Aoba, Sendai 980-8578, Japan} 

\author[0000-0001-7146-4687]{Yen-Ting Lin}
\affiliation{Academia Sinica Institute of Astronomy and Astrophysics, 11F of Astronomy-Mathematics Building, AS/NTU, No.1, Section 4, Roosevelt Road, Taipei 10617, Taiwan}

\author[0000-0002-3531-7863]{Yoshiki Toba}
\affiliation{Department of Physical Sciences, Ritsumeikan University, 1-1-1 Noji-higashi, Kusatsu, Shiga 525-8577, Japan}
\affiliation{Academia Sinica Institute of Astronomy and Astrophysics, 11F of Astronomy-Mathematics Building, AS/NTU, No.1, Section 4, Roosevelt Road, Taipei 10617, Taiwan}
\affiliation{Research Center for Space and Cosmic Evolution, Ehime University, 2-5 Bunkyo-cho, Matsuyama, Ehime 790-8577, Japan}

\author{Sakiko Obuchi}
\affiliation{Department of Physics, Graduate School of Advanced Science and Engineering, Faculty of Science and Engineering, Waseda University, 3-4-1, Okubo, Shinjuku, Tokyo 169-8555, Japan}

\author[0000-0002-5956-8018]{Itsna Khoirul Fitriana}
\email{}
\affiliation{Astronomy Research Group, Institut Teknologi Bandung, Jl. Ganesha No. 10 Bandung 40132, Indonesia}
\affiliation{National Astronomical Observatory of Japan, 2-21-1, Osawa, Mitaka, Tokyo 181-8588, Japan}

\begin{abstract}
We present a wide-area and deep optical identification catalog for radio sources based on the Very Large Array Sky Survey (VLASS) Epoch 2 catalog at 3~GHz.
Optical counterparts are identified using the final-year internal processing of the Hyper Suprime-Cam Subaru Strategic Program (HSC-SSP) Wide layer (DR S23B), which provides deep imaging over $\sim1200$~deg$^{2}$ in $grizy$ filters with an $i$-band depth of $i_\mathrm{AB}\simeq26$.
Starting from a $1\farcs0$ match between VLASS and HSC, we construct a primary catalog of 22,773 sources by requiring a signal-to-noise ratio of $>5$ in at least one HSC band.  
We further provide nearest-neighbor associations to the Faint Images of the Radio Sky at Twenty Centimeters (FIRST; 1.4\,GHz) and the Low Frequency Array Two-metre Sky Survey (LoTSS; 144\,MHz) within $2\farcs5$, resulting in 18,444 FIRST-matched sources, 16,167 LoTSS-matched sources, and a 14,206-source subset matched to both surveys. 
The catalog contains approximately six times more sources than the first Wide-field Exploration of Radio-detected active Galactic nuclei with Subaru (WERGS) catalog based on the early HSC-SSP S16B data and positional cross-matching with FIRST \citep{Yamashita2018WERGS1}, while the higher VLASS resolution enables more precise optical associations. 
Compared to the Ultraviolet Near-Infrared Optical Northern Survey (UNIONS)-based VLASS identifications \citep{Zhong2025UNVEIL1}, the deeper HSC imaging improves sensitivity to optically faint and morphologically resolved hosts at $z\gtrsim1$. 
Our catalog preferentially highlights host-dominated active galactic nucleus candidates, potentially including a substantial fraction of obscured systems. 
\end{abstract} 


\section{Introduction}\label{sec:intro}
Supermassive black holes (SMBHs) in nearby galaxies exhibit tight empirical correlations with the properties of their host bulges. 
The black hole mass and bulge velocity dispersion ($M_{\rm BH}$--$\sigma_\ast$) relation provides one of the most widely used links between SMBH growth and galaxy evolution
\citep[e.g.,][]{FerrareseMerritt2000,Gebhardt2000,Tremaine2002}.
Despite its apparent tightness in the local universe, the physical origin of the relation remains debated.
A broad class of models attributes it to self-regulated SMBH growth,
where energy and/or momentum injection from active galactic nuclei (AGN)
couples to the surrounding gas and limits further accretion and star formation
\citep[e.g.,][]{SilkRees1998,King2003}.

A key observational pathway to constrain such feedback scenarios is offered by radio 
galaxies \footnote{ In general, the term ``radio galaxies'' often refers to edge-on (type-2) radio AGN whose host galaxies are visible in the optical band \citep[e.g.,][]{ant93,UrryPadovani1995}. 
In this study, however, we adopt a broader definition and use the term to refer to all radio-detected AGN, including both type-1 and type-2 systems, as well as radio-loud quasars. }. 
Powerful radio jets can inject kinetic energy into the circumgalactic and interstellar media,
heating or expelling gas and thereby suppressing (or in some circumstances redistributing) star formation on galactic scales \citep{McNamaraNulsen2007,Fabian2012,HeckmanBest2014,Morganti2017,Heckman2024}.
This ``radio-mode'' (kinetic) feedback is frequently invoked to explain the quenching of massive galaxies and the maintenance of hot  gaseous halos, and it is commonly implemented in semi-analytic and hydrodynamical models to reproduce the observed abundance and colours of massive galaxies and the bright end of luminosity and stellar-mass functions \citep[e.g.,][]{Croton2006,Bower2006,Vogelsberger2014,Schaye2015}.
If radio-mode feedback is indeed a major channel by which SMBHs influence their hosts,
statistical samples of radio galaxies with well-characterized host properties
can provide essential constraints on the establishment of black hole--galaxy scaling relations. 

The search for radio galaxies has made major strides with the advent of wide-area radio surveys covering a broad range of frequencies and angular resolutions. 
At GHz frequencies, the Faint Images of the Radio Sky at Twenty Centimeters (FIRST) survey provides imaging over a large area with an angular resolution (synthesized-beam FWHM) of $\sim6^{\prime\prime}$ \citep[][]{Becker1995,Helfand2015}, while the Very Large Array Sky Survey (VLASS) delivers higher angular resolution of $\sim2^{\prime\prime}$ at 2--4\,GHz \citep[][]{Lacy2020}. 
At low frequencies, LOw-Frequency ARray (LOFAR) Two-metre Sky Survey \citep[LoTSS;][]{Shimwell17,Shimwell2022,Shimwell2026} provides sensitive 120--168\,MHz imaging at $\sim6^{\prime\prime}$ angular resolution (with additional 20$^{\prime\prime}$ images for diffuse emission), enabling robust spectral-index measurements and improved identification of extended jet/lobe systems when combined with GHz data.

Radio galaxy samples were often built from bright, low-frequency flux-limited catalogues such as 3CRR \citep{Laing1983},
and high-redshift radio galaxies were efficiently targeted via the ultra-steep spectrum (USS) technique,
motivated by the empirical spectral-index--redshift correlation \citep{Roettgering1994,DeBreuck2000,mil08}.
Modern multi-frequency surveys now allow these classical approaches to be revisited with far larger, more homogeneous samples,
and to be coupled directly to well-characterized host-galaxy properties from optical/near-infrared imaging.
However, selections based on spectral steepness remain inherently biased:
samples constructed from steep-spectrum criteria are preferentially weighted toward lobe-dominated sources
and can therefore yield an incomplete view of the overall radio-AGN population.

An alternative route to identifying radio-galaxy hosts without spectral-slope pre-selection has long been large-area radio--optical cross-matching between radio surveys (e.g., FIRST/NRAO VLA Sky Survey (NVSS)) and wide-field optical imaging such as Sloan Digital Sky Survey (SDSS)  \citep[e.g.,][]{Ivezic2002,Best2005,Kimball2008,Lin2018}.
\citet{Ivezic2002} analyzed $\sim3\times10^{4}$ FIRST sources with SDSS counterparts over 1,230~deg$^{2}$, and \citet{Best2005} constructed a catalog of 2,712 radio-luminous galaxies by matching the SDSS DR2 main spectroscopic sample to NVSS/FIRST over $\approx$2,627~deg$^{2}$.
\citet{Kimball2008} further examined the $\sim$3,000~deg$^{2}$ overlap region containing $\sim$140,000 NVSS--FIRST sources, of which $\sim$64{,}000 have SDSS detections. 
More recently, \citet{Lin2018} constructed a catalog of $\sim$2300 low-radio-luminosity active galaxies based on FIRST/NVSS and SDSS data to investigate the origin of nuclear activity in radio galaxies. 
These studies established statistical samples of optically identified radio sources and radio-loud AGN, 
but were ultimately limited by the relatively shallow depth of SDSS ($r\lesssim 22$--23\,mag), leaving many faint and/or high-redshift radio-source hosts unidentified. 
Deeper optical imaging is therefore essential to extend such identifications to the faint end and to the rarest high-redshift population. 

Recently, the Ultraviolet Near-Infrared Optical Northern Survey \citep[UNIONS; ][]{gwy25} provides deep, wide-area optical imaging over a large fraction of the northern sky. 
Building on this dataset, the UNIONS Optical Identifications for VLASS Radio Sources in the Euclid Sky (UNVEIL) program has recently delivered a wide-area optical identification of VLASS sources, constructing a catalog of 146,212 radio AGN candidates down to $r\simeq24.5$ over $\sim$4,200~deg$^2$ \citep{Zhong2025UNVEIL1}. Early science from this catalog has already led to the discovery of a radio quasar with a peculiar spectral shape, which has not been reported before \citep{Zhong2026BBQSORS}.
In addition, the UNIONS--VLASS footprint substantially overlaps the Euclid wide survey region \citep{EuclidCollaboration2022EWS}, offering a complementary legacy dataset for statistical studies of radio-loud AGNs and their host galaxies in the Euclid era \citep{Zhong2025UNVEIL1}. 

The Wide and deep Exploration of Radio-detected active Galactic nuclei with Subaru/HSC \citep[WERGS;][]{Yamashita2018WERGS1,Toba2019WERGS2,Yamashita2020WERGS3,Ichikawa2021WERGS4,Uchiyama2022WERGS6,Uchiyama2022WERGS7,Uchiyama2022WERGS9,Ichikawa2023eFEDS_WERGS,Yamamoto2025WERGS10}
was initiated to construct a radio galaxy sample without relying on spectral-slope pre-selection, by leveraging the depth and image quality of the Hyper Suprime-Cam (HSC) Subaru Strategic Program (HSC-SSP; \citealt{Aihara2022HSC_DR3,Miyazaki2018}).
With multi-band imaging reaching $i \sim 26$\,mag and excellent image quality over a wide area, HSC-SSP enables reliable identification of faint optical counterparts to radio sources and robust measurements of their host-galaxy properties, well beyond what was possible with other wide-field optical surveys such as SDSS and UNIONS.
In its first catalog implementation (WERGS~I; \citealt{Yamashita2018WERGS1}), radio sources from the FIRST 1.4\,GHz survey were cross-matched with optical counterparts in the early HSC-SSP Wide effective area (DR~S16A; 154~deg$^{2}$), establishing a baseline optically identified radio-source sample in a manner that does not require spectral-slope pre-selection.
While this approach demonstrated the power of deep HSC imaging for counterpart identification, the limited HSC area available at the time and the relatively modest angular resolution of FIRST (FWHM $\sim6^{\prime\prime}$)  constrained the counterpart sample size and the recovery of the rarest high-redshift systems.
Nevertheless, this capability has proven essential for uncovering rare and/or distant systems, including dust-obscured galaxies with prominent jet signatures \citep{fuk25}, the discovery of a radio (and X-ray luminous) quasar at $z=3.4$ undergoing super-Eddington accretion \citep{obu26}, as well as the identification of a $z=4.72$ radio galaxy, the highest-$z$ type-2 radio AGN \citep{Yamashita2020WERGS3}.

\begin{figure*}
  \centering
  \includegraphics[width=\textwidth]{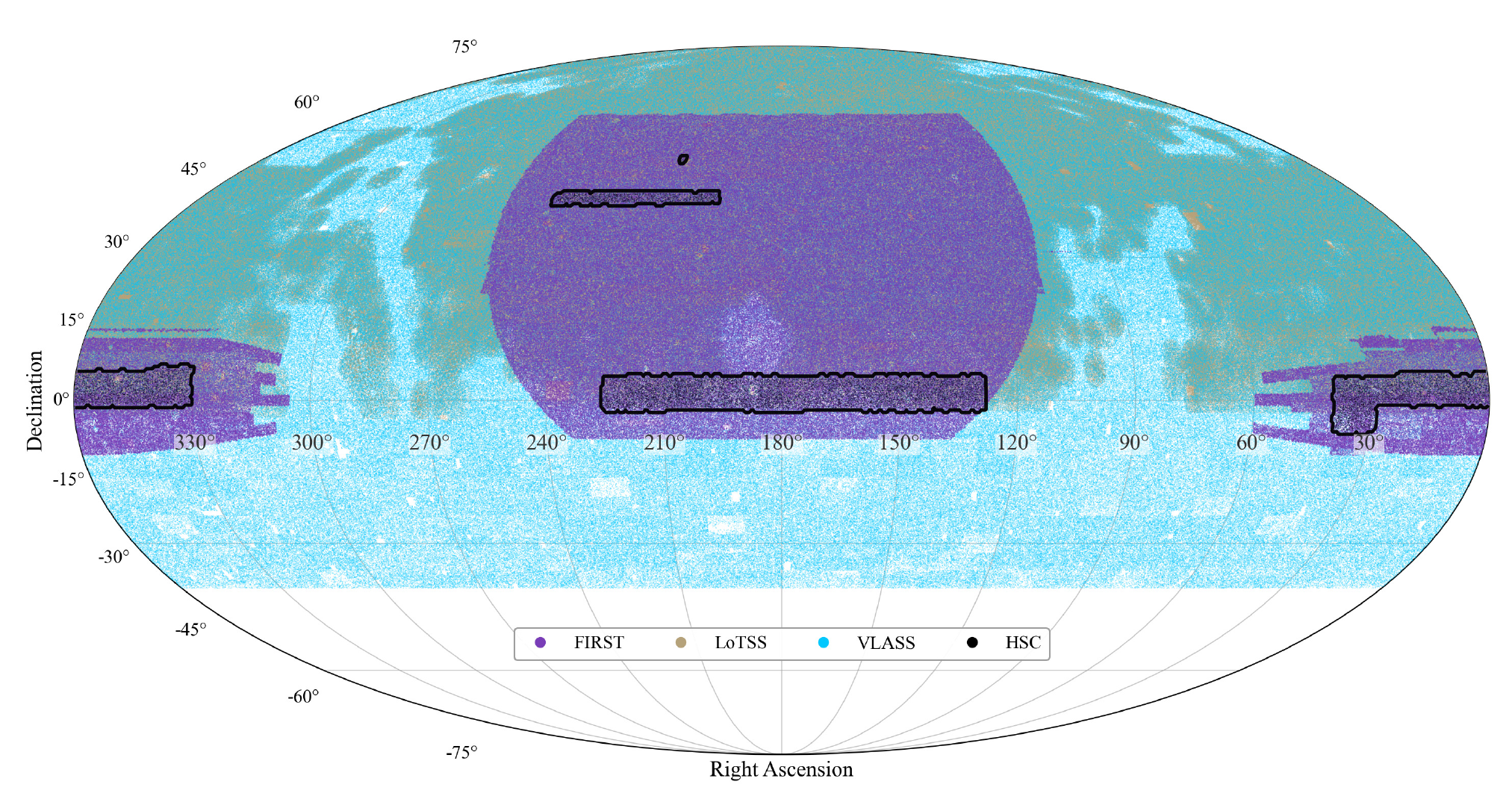}
  \caption{
  Sky footprints of the surveys used in this work, shown in equatorial coordinates.
  Dots indicate the sky positions of cataloged sources in each survey: VLASS Epoch~2 (cyan), FIRST (purple), LoTSS DR3 (khaki), and the HSC-SSP Wide-layer matched sample (black). The sources detected both in VLASS Epoch~2 and LoTSS DR3 are shown in green because of the combined color of the both. The black outline highlights the approximate footprint of the HSC-SSP Wide layer. 
  }
  \label{fig:footprint}
\end{figure*}

In this paper, we present the optical-identification results based on the final processing of the HSC-SSP Wide layer, covering the footprint of $\sim$1200~deg$^{2}$ with multi-band imaging reaching $i\simeq26$\,mag. 
We cross-match the HSC-SSP optical sources with major radio surveys spanning $0.1-3$ GHz frequencies 
(such as LOFAR/LoTSS, FIRST, and VLASS) to build a homogeneous, multi-frequency radio source sample with robust optical host identifications.
While the HSC-SSP footprint is smaller than that of UNIONS, its substantially greater depth and image quality enable us to identify much fainter host galaxies and to characterize their properties in a uniform manner. 
This depth is particularly important for capturing extreme radio-loud systems whose optical hosts are too faint to be recovered in shallower wide-area imaging \citep[][]{Ichikawa2021WERGS4}, thereby extending the dynamic range of radio loudness and host-galaxy properties accessible for statistical studies. 
The resulting catalog provides a deep optical foundation for downstream analyses, including redshift estimation, radio-loudness measurements, and host-galaxy characterization, and is intended as a legacy dataset for studies of radio-mode AGN feedback and galaxy evolution.

This paper presents the catalog construction, quality assessment, and basic properties
of our cross-matched sample. 
Throughout this paper, we assume a flat $\Lambda$CDM cosmology with
$H_0 = 67.4~{\rm km~s^{-1}~Mpc^{-1}}$, $\Omega_{\rm m}=0.315$, and $\Omega_\Lambda=0.685$ \citep[][]{Planck2018VI}, and we use AB magnitudes for photometry.

\section{DATA}\label{sec:data}

In this section, we briefly describe the four surveys used to construct our multi-wavelength radio-galaxy catalog:
VLASS (3\,GHz), FIRST (1.4\,GHz), LoTSS (144\,MHz), and the Subaru HSC-SSP Wide-layer optical imaging.
Figure~\ref{fig:footprint} summarizes the sky coverage of these surveys and their overlaps.

\subsection{VLASS}
VLASS is a $2-4$ GHz synoptic radio survey covering the sky north of declination $\delta \ge -40^\circ$ ($\sim$3~million sources in $\sim34,000$ deg$^2$) in three epochs (Epoch 1: 2017–2019; Epoch 2: 2020–2022; Epoch 3: 2023–2025), each providing a full pass of the same footprint separated by $\sim32$ months \citep[][and see also Figure~\ref{fig:footprint}]{Lacy2020}.  
In this paper, we adopt the VLASS Epoch~2 Single-Epoch (SE) images and component catalog,
generated from the Epoch~2 observations and released by the Canadian Initiative for Radio Astronomy Data Analysis 
\citep[CIRADA;][]{Gordon2023CIRADAVLASSUserGuide,Lacy2022VLASSMemo17}. 
The Epoch 2 SE products provide a uniform, publicly released reference data set for this work. 
We do not use the Epoch 1 imaging, which is known to be affected by issues such as antenna pointing errors that impact image fidelity and are addressed in later processing/re-observations, and we also avoid Epoch~3 products because, at the time of our analysis, fully validated SE images/component catalogs were not yet available in the same mature release state as Epoch 2 \citep[e.g., some Epoch~3 products are distributed as interim/quick-look releases; ][]{Lacy2022VLASSMemo17}.

The SE products correspond to images/catalogs generated from individual observing epochs, prior to stacking across multiple epochs. 
The images have a $\sim2.2-3.0$ arcsec synthesized beam, $0.6$ arcsec pixels, and median rms $\sim120-170$ $\mu$Jy determined from blank regions. 
Source finding for the VLASS imaging products is typically carried out with a flood-fill–based algorithm (e.g., PyBDSF\footnote{\texttt{PyBDSF} (Python Blob Detection and Source Finder) is a widely used source-finding software for radio-interferometric images that detects emission islands and fits Gaussian components to construct catalogs \citep{pybdsf_ascl}.}), using a $\sim5\sigma$ peak-detection threshold and a lower $\sim3\sigma$ threshold to define contiguous emission ``islands", where $\sigma$ is the local rms noise. 
The SE catalog lists sources detected at peak significance with $\ge 5\sigma$, corresponding to a representative limit of $\sim0.6$--$0.9$~mJy for the median rms (with spatially varying depth across the footprint). 
We do not impose an additional uniform flux-density cut beyond the catalog-level detection threshold.
The astrometric uncertainty of VLASS SE sources can be approximated as $\sim0.5$ arcsec for sources with a signal-to-noise ratio (S/N) of $\sim5$, improving to $<0.3$ arcsec for bright sources (see the VLASS SE Continuum Users Guide\footnote{\url{https://science.nrao.edu/vlass/vlass-se-continuum-users-guide}}).

\subsection{FIRST}
The FIRST survey is a 1.4 GHz radio continuum survey carried out with the Very Large Array (VLA) in B-configuration \citep[][]{Becker1995,Helfand2015}.
It covers $\sim10{,}000$ square degrees primarily in the North and South Galactic Caps, overlapping well with the SDSS  footprint.
The final catalog contains nearly one million radio sources detected at a typical rms noise level of $\sim0.15$ mJy beam$^{-1}$, with a synthesized beam of $\sim6$ arcsec FWHM and $1.8$ arcsec pixels. 
The FIRST catalog includes only sources with peak flux densities exceeding five times the local rms noise, $F_{\rm peak} \ge 5\,{\rm rms}$. 
The astrometric accuracy of FIRST sources is better than $0.5$ arcsec for sources brighter than a few mJy and degrades to $\lesssim1$ arcsec near the detection threshold, sufficient for robust cross-matching with optical and infrared surveys.

\subsection{LoFAR}
The LOFAR/LoTSS is a low-frequency (120--168 MHz) radio continuum survey of the northern sky with LOFAR \citep[][]{Shimwell2026}.
Its third data release (DR3) covers 88\% of the northern sky at a central frequency of 144 MHz, delivering $6''$ and $20''$ resolution images.
The $6''$ resolution mosaics have a median rms sensitivity of $92~\mu{\rm Jy~beam}^{-1}$ and an astrometric accuracy of $0.24''$, and the associated source catalogue contains 13,664,379 radio sources.
In this paper, we adopt the LoTSS-DR3 images and associated source catalogues provided by the LOFAR Surveys team.\footnote{\url{https://lofar-surveys.org/dr3.html}}


\subsection{Subaru HSC-SSP}
The Subaru HSC-SSP is a deep-and-wide imaging survey carried out with the HSC on the Subaru Telescope, which provides a $1.5^{\circ}$-diameter field of view \citep{Aihara2018}.
In this study, we use the Wide-layer data from the HSC-SSP final-year internal dataset (DR S23B), which will correspond to the forthcoming final public data release \citep[PDR4; ][]{Oguri2025S23B,HSCSSPDataRelease}.
The Wide layer covers $\sim1200$~deg$^2$ in the five broad bands $g$, $r$, $i$, $z$, and $y$, 
with median $i$-band seeing $\sim0\farcs6$ \citep{Aihara2018,HSCSSPSurvey}.
The Wide-layer footprint consists of three large contiguous regions: two equatorial stripes (spring and autumn) and the HectoMAP region at $\mathrm{Dec}\sim40$~deg—plus the AEGIS field, which is a single HSC pointing observed to the Wide-layer depth \citep{Aihara2018}.
The survey design and the filter information are given in \citet{Aihara2018} and \citet{Kawanomoto2018}, respectively. 
\citet{Komiyama18} should be the reference for the camera system and the CCD dewar designs. 
Among the HSC-SSP observations, the on-site quality assurance system for the HSC \citep[OSQAH; ][]{Furusawa18} was used in order to 
provide real-time feedback to the observations. 
The dedicated pipeline hscPipe \citep[][]{Bosch2018}, which is a modified version of the Legacy Survey of Space and Time software stack \citep[][]{Ivezic19, Juric17, Bosch19}, was used for data reduction.  
Photometric calibration is tied to the Pan-STARRS1 system \citep[][]{Cambers16, Schlafly12, Tonry12, Magnier13}, and astrometric calibration is performed using Gaia DR2 through Jointcal \citep[][]{Gaia2018}. 


\section{Sample selection}

\begin{figure*}
\begin{center}
\includegraphics[width=0.65\linewidth]{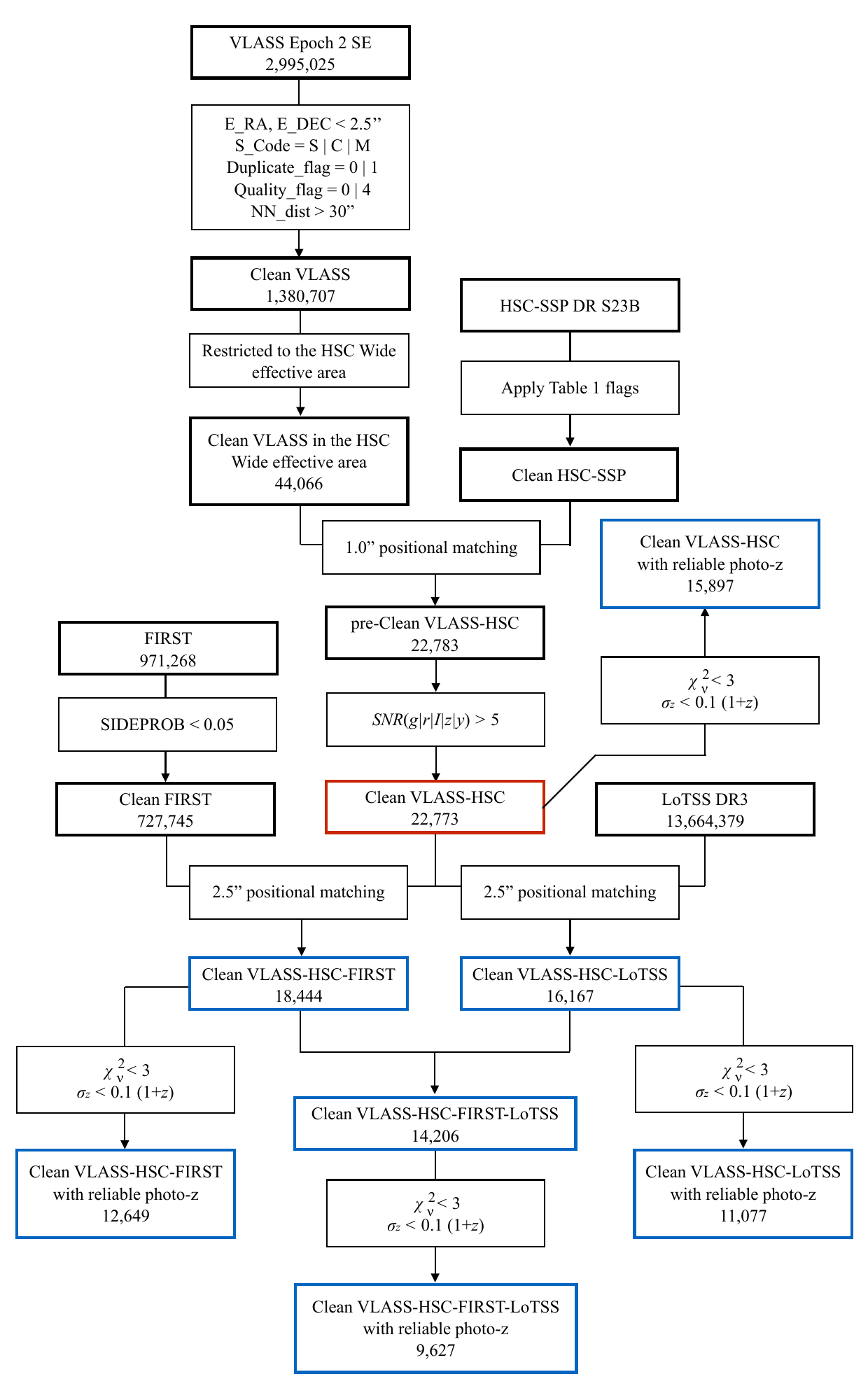}
\end{center}
\caption{
Flowchart summarizing the construction of our multi-wavelength samples.
Starting from the VLASS Epoch~2 SE component catalog and the HSC-SSP DR~S23B Wide-layer photometric catalog, we apply a series of quality cuts to define the Clean VLASS and Clean HSC-SSP samples.
We then perform a $1\farcs0$ positional match between Clean VLASS and Clean HSC-SSP to construct the Clean VLASS--HSC catalog ($N=22{,}773$; red-outlined box), which serves as the parent sample for subsequent cross-matching.
From this parent catalog, we cross-match to the Clean FIRST source catalog (selected with \texttt{SIDEPROB}$<0.05$) and to the LoTSS DR3 catalog within $2\farcs5$, producing the Clean VLASS--HSC--FIRST ($N=18{,}444$), Clean VLASS--HSC--LoTSS ($N=16{,}167$), and the intersection sample with counterparts in both FIRST and LoTSS, Clean VLASS--HSC--FIRST--LoTSS ($N=14{,}206$).
Blue-outlined boxes denote auxiliary subsamples with reliable photometric redshifts, selected by $\chi^2_\nu<3$ and $\sigma_z<0.1(1+z)$, yielding $N=15{,}897$ (Clean VLASS--HSC with reliable photo-$z$), $N=12{,}649$ (Clean VLASS--HSC--FIRST with reliable photo-$z$), $N=11{,}077$ (Clean VLASS--HSC--LoTSS), and $N=9,627$ (Clean VLASS--HSC--FIRST--LoTSS with reliable photo-$z$).
}\label{flow}
\end{figure*}

Following the workflow illustrated in Figure \ref{flow}, we construct our radio galaxy sample step by step. 
First, we begin from the VLASS Epoch~2 SE catalog, adopting it as the parent list of radio detections.
The parent catalog contains a total of 2,995,025 entries.
We then define a quality-controlled subset (hereafter the Clean VLASS catalog) by applying catalog-level cuts,
following the procedure of \citet{Zhong2025UNVEIL1}, to remove obvious imaging pathologies and non-astrophysical artifacts
while retaining bona fide radio sources. 
Specifically:
(i) we require well-constrained astrometry,
E\_RA $< 2.5^{\prime\prime}$ and E\_DEC $< 2.5^{\prime\prime}$,
where E\_RA and E\_DEC are the quoted 1$\sigma$ positional uncertainties in right ascension and declination, respectively;
(ii) we accept only standard PyBDSF source classes,
S\_Code $\in{\mathrm{S,M,C}}$, where $\mathrm{S}$ indicates a source fit by a single Gaussian, $\mathrm{M}$ a source requiring multiple Gaussians, and $\mathrm{C}$ a source fit by a single Gaussian but lying in the same flux island as another source (i.e., a blended/complex island);
(iii) we restrict to entries with acceptable duplication and detection-quality flags,
Duplicate\_flag $\in{0,1}$ and Quality\_flag $\in{0,4}$,
where Duplicate\_flag identifies repeated detections of the same astrophysical component in overlapping images (we exclude explicit duplicates), and Quality\_flag encodes detection reliability; Quality\_flag $=0$ corresponds to unflagged detections, while Quality\_flag $=4$ denotes cases where the integrated flux density is formally smaller than the peak brightness but the detection is otherwise acceptable.
Finally, we impose NN\_dist $>30^{\prime\prime}$, where NN\_dist is the angular separation to the nearest other fitted catalog component after accounting for duplicates, in order to reduce obvious blending/confusion in subsequent cross-matching and morphology measurements. 
We note that the NN\_dist $>30^{\prime\prime}$ criterion may preferentially exclude highly extended or multi-component radio galaxies, including some classical double-lobed radio galaxies. 
The resulting catalog is therefore optimized for robust optical counterpart identification and statistical analyses of predominantly compact radio sources, rather than completeness for highly extended radio morphologies. 
As a result, we obtain quality-controlled subset of 1,380,707 objects as the Clean VLASS catalog. 

\newcommand{\reqbox}[1]{\parbox[t]{0.32\textwidth}{\raggedright\ttfamily #1}}

\begin{deluxetable*}{lll}
\tablecaption{HSC-SSP quality cuts for optical counterparts\label{tab:hsc_clean_flags}}
\tabletypesize{\scriptsize}
\tablehead{
\colhead{Category} & \colhead{Requirement} & \colhead{Purpose}
}
\startdata
Primary detection
  & \reqbox{main.isprimary = `t'}
  & Keep primary objects and avoid duplicate deblends. \\
Edge / interpolation
  & \reqbox{%
    main.\{g,r,i,z,y\}\_pixelflags\_edge = `f'\\
    main.\{g,r,i,z,y\}\_pixelflags\_interpolatedcenter = `f'
  }
  & Reject edge-affected measurements and center-interpolation failures. \\
\parbox[t]{0.18\textwidth}{\raggedright
Saturation /\\
cosmic ray /\\
bad pixels}
  & \reqbox{%
    main.\{g,r,i,z,y\}\_pixelflags\_saturatedcenter = `f'\\
    main.\{g,r,i,z,y\}\_pixelflags\_crcenter = `f'\\
    main.\{g,r,i,z,y\}\_pixelflags\_bad = `f'
  }
  & Ensure robust photometry at the source center. \\
Centroiding (meas2)
  & \reqbox{meas2.\{g,r,i,z,y\}\_sdsscentroid\_flag = `f'}
  & Require well-behaved centroid measurements. \\
Bright-star masks
  & \reqbox{%
    masks.\{g,r,i,z,y\}\_mask\_brightstar\_halo = `f'\\
    masks.\{g,r,i,z,y\}\_mask\_brightstar\_ghost = `f'\\
    masks.\{g,r,i,z,y\}\_mask\_brightstar\_blooming = `f'\\
    masks.y\_mask\_brightstar\_channel\_stop = `f'
  }
  & Exclude regions contaminated by bright-star artifacts. \\
Coverage (inputcount)
  & \reqbox{%
    main.g\_inputcount\_value $\ge 3$ main.r\_inputcount\_value $\ge 3$\\
    main.i\_inputcount\_value $\ge 5$ main.z\_inputcount\_value $\ge 5$ main.y\_inputcount\_value $\ge 5$
  }
  & Require sufficient number of contributing visits for stable measurements. \\
\enddata
\tablecomments{Cuts are applied to the HSC forced-photometry tables in the Wide layer (field \texttt{s23b\_wide} in this work), as implemented in our SQL queries.}
\end{deluxetable*}

Second, on the HSC side, we enforce a conservative ``clean'' selection to ensure robust photometry and astrometry for the optical counterparts, as summarized in  Table~\ref{tab:hsc_clean_flags}. 
In brief, we require primary detections, exclude measurements affected by image edges or problematic central pixels (e.g., interpolation, saturation, cosmic rays, or bad pixels), and ensure reliable centroid measurements in all bands. 
We also mask regions contaminated by bright-star artifacts and require a sufficient number of contributing exposures in each band to guarantee stable measurements. 
After applying these criteria, we retrieve multi-band photometry (e.g., \texttt{cmodel} and PSF fluxes) and ancillary quantities such as Galactic extinction coefficients and photometric-redshift \citep[Mizuki;][]{Tanaka18} estimates when available. 
We also attach spectroscopic redshifts from the HSC-SSP spec-$z$ compilation when available.

Of the 1,380,707 sources in the Clean VLASS catalog, 62,867 fall within the nominal HSC--SSP Wide footprint. After applying the HSC-side masking and quality cuts used for the clean optical sample (except for the \texttt{sdsscentroid} requirement; Appendix~B), 44,066 remain within the effective HSC footprint.

We then cross-match the Clean VLASS sample and the Clean HSC-SSP sample using a $1\farcs0$ nearest-neighbor positional match.
This matching radius choice balances completeness and reliability, and guarantees a contamination rate $\lesssim10$\% under our source-density and astrometric assumptions (see Appendix~A for more details). 
This yields an initial cross-matched catalog, which we refer to as the pre-Clean VLASS--HSC sample, containing 22,783 sources (Figure~\ref{flow}).
To ensure that each matched object has a reliable optical detection, we further require S/N of $>5$ in at least one HSC band, evaluated from the \texttt{cmodel} magnitude uncertainties.
After applying this requirement, we obtain our Clean VLASS--HSC catalog containing 22,773 sources, of which 5,211 have spectroscopic redshifts.
Relaxing the per-band photometric signal-to-noise threshold to S/N$>4$ increases the sample size only marginally to 22,781 sources, indicating that the catalog is not strongly sensitive to the adopted threshold. 


Next, we construct a FIRST-matched subsample by requiring a counterpart in the FIRST 1.4\,GHz catalog for sources in the Clean VLASS--HSC parent sample.
We retain reliable FIRST detections by applying \texttt{SIDEPROB}$<0.05$ (i.e., a low sidelobe-contamination probability; \citealt{Helfand2015}), and then perform a nearest-neighbor positional cross-match.
We adopt a matching radius of $2\farcs5$, consistent with \citet{Zhong2025UNVEIL1}.
The parent FIRST catalog contains 971,268 sources, and the \texttt{SIDEPROB}$<0.05$ cut yields 727,745 objects, which we refer to as the Clean FIRST sample.
Requiring a Clean FIRST counterpart within $2\farcs5$ defines the Clean VLASS--HSC--FIRST sample, containing 18,444 sources (81\% of the 22,773 Clean VLASS--HSC objects; Figure~\ref{flow}), including 4,240 with spectroscopic redshifts. 

We also construct a LoTSS-matched subsample by requiring a counterpart in the LoTSS DR3 source catalog, using the same nearest-neighbor matching scheme and a radius of $2\farcs5$ \citep[cf.][]{Zhong2025UNVEIL1}.
This counterpart requirement defines the Clean VLASS--HSC--LoTSS sample, containing 16,167 sources (71.1\% of the parent sample; Figure~\ref{flow}), including 3,478 sources with spectroscopic redshifts. 

In addition, we define the intersection sample that has counterparts in both Clean FIRST and LoTSS DR3 within $2\farcs5$.
This yields the Clean VLASS--HSC--FIRST--LoTSS sample with $N=14{,}206$ (Figure~\ref{flow}), including 3,042 with spectroscopic redshifts. 

Finally, to enable analyses that require robust redshift information, we define ``with reliable photo-$z$'' subsamples for each of the above subsamples (blue-outlined boxes in Figure~\ref{flow}).
Specifically, we apply the photometric-redshift quality cut $\chi^2_\nu<3$ and $\sigma_z<0.1(1+z)$, where $\chi^2_\nu$ is the reduced $\chi^2$ of the photo-$z$ fit and $\sigma_z$ is the quoted photo-$z$ uncertainty.
This yields the Clean VLASS--HSC with reliable photo-$z$ sample ($N=15{,}897$), the Clean VLASS--HSC--FIRST with reliable photo-$z$ sample ($N=12{,}649$), the Clean VLASS--HSC--LoTSS with reliable photo-$z$ sample ($N=11{,}077$), and the Clean VLASS--HSC--FIRST--LoTSS with reliable photo-$z$ sample ($N=9,627$) (Figure~\ref{flow}).
For reference, the numbers of sources with available spectroscopic redshifts in these subsamples are 3,797, 3,020, 2,461, and 2,121, respectively.


\section{Catalog Description}\label{sec:catalog}

We provide our parent multi-wavelength catalog as a machine-readable table. 
The full table contains 22,773 entries and corresponds to the Clean VLASS--HSC parent catalog (red-outlined box in Figure~\ref{flow}; Section~3). 
It also includes all columns needed to reproduce the blue-outlined subsamples in Figure~\ref{flow}. 
For each source, the table includes (i) VLASS 3\,GHz radio properties, (ii) HSC-SSP optical photometry and quality-screened counterpart information, (iii) photometric-redshift and stellar-mass estimates from the Mizuki catalog \citep{Tanaka18} when available, (iv) nearest-neighbor match information to FIRST (1.4\,GHz) and LoTSS (144\,MHz) within $2\farcs5$ (including counterpart identifiers and separations), and (v) available spectroscopic redshifts from the HSC external spec-$z$ compilation. 
A subset of key columns is summarized in Table~\ref{tab:catalog_columns}.

\newcommand{\cname}[1]{\texttt{#1}}
\newcommand{\cdesc}[1]{\parbox[t]{0.62\textwidth}{\raggedright #1}}

\begin{deluxetable*}{lll}
\tablecaption{Summary of key columns in the released catalog\label{tab:catalog_columns}}
\tabletypesize{\scriptsize}
\tablehead{
\colhead{Column} & \colhead{Unit} & \colhead{Description}
}
\startdata
\multicolumn{3}{l}{\textit{Identifiers and positions}}\\ \hline
\cname{Component\_name} & \nodata & VLASS component identifier.\\
\cname{RA}, \cname{DEC} & deg & VLASS position (ICRS).\\
\cname{E\_RA}, \cname{E\_DEC} & deg & VLASS positional uncertainties.\\
\cname{hsc\_object\_id} & \nodata & Matched HSC object identifier.\\
\cname{hsc\_ra}, \cname{hsc\_dec} & deg & HSC counterpart position.\\
\cname{hsc\_match\_dist\_arcsec} & arcsec & VLASS--HSC separation.\\ \hline
\multicolumn{3}{l}{\textit{VLASS 3\,GHz radio measurements}}\\ \hline
\cname{Total\_flux}, \cname{E\_Total\_flux} & mJy & Integrated 3\,GHz flux density and uncertainty.\\
\cname{Peak\_flux}, \cname{E\_Peak\_flux} & mJy beam$^{-1}$ & Peak 3\,GHz flux density and uncertainty.\\ \hline
\multicolumn{3}{l}{\textit{HSC photometry and Mizuki photo-$z$}}\\ \hline
\cname{hsc\_\{g,r,i,z,y\}\_cmodel\_mag} & mag & HSC \texttt{cmodel} magnitudes in $grizy$.\\
\cname{hsc\_\{g,r,i,z,y\}\_cmodel\_magerr} & mag & Uncertainties of \texttt{cmodel} magnitudes.\\
\cname{hsc\_\{g,r,i,z,y\}\_psfflux\_mag} & mag & PSF magnitudes in $grizy$.\\
\cname{hsc\_\{g,r,i,z,y\}\_psfflux\_magerr} & mag & Uncertainties of PSF magnitudes.\\
\cname{hsc\_a\_\{g,r,i,z,y\}} & mag & Galactic extinction coefficients.\\
\cname{hsc\_photoz\_best} & \nodata & Mizuki photo-$z$ estimate \citep{Tanaka18}.\\
\cname{hsc\_photoz\_std\_best} & \nodata & Photo-$z$ uncertainty proxy used for error propagation.\\
\cname{hsc\_specz} & \nodata & Spectroscopic redshift compiled for HSC counterparts (when available).\\
\cname{hsc\_specz\_err} & \nodata & Uncertainty of \cname{hsc\_specz}.\\
\cname{hsc\_stellar\_mass} & $M_\odot$ & Mizuki stellar mass estimate.\\
\cname{hsc\_stellar\_mass\_err68\_min/max} & $M_\odot$ & 68\% confidence bounds for stellar mass.\\ \hline
\multicolumn{3}{l}{\textit{FIRST and LoTSS match information}}\\ \hline
\cname{first\_SIDEPROB} & \nodata & FIRST sidelobe-contamination probability \citep{Helfand2015}.\\
\cname{first\_match\_dist\_arcsec} & arcsec & Nearest-neighbor separation to FIRST.\\
\cname{first\_FINT} & mJy &  FIRST integrated flux density.\\
\cname{first\_FPEAK} &  mJy beam$^{-1}$ &  FIRST peak flux density.\\
\cname{first\_RMS} &  mJy beam$^{-1}$ &  Local RMS noise of the FIRST image at the source position. \\
\cname{lotss\_match\_dist\_arcsec} & arcsec & Nearest-neighbor separation to LoTSS.\\
\cname{lotss\_Total\_flux}, \cname{lotss\_E\_Total\_flux} & mJy & LoTSS integrated 144\,MHz flux density and uncertainty.\\
\cname{lotss\_Peak\_flux}, \cname{lotss\_E\_Peak\_flux} & mJy beam$^{-1}$ & LoTSS peak flux density and uncertainty.\\ \hline
\multicolumn{3}{l}{\textit{Derived redshift and optical morphology}}\\ \hline
\cname{z\_best} & \nodata & Best-available redshift (priority: \cname{hsc\_specz} if available; otherwise \cname{hsc\_photoz\_best}).\\
\cname{z\_err\_best} & \nodata & Redshift uncertainty (priority: \cname{hsc\_specz\_err}, else \cname{hsc\_photoz\_std\_best}).\\
\cname{delta\_i} & mag & $i$-band extendedness $\Delta i \equiv i_{\rm PSF}-i_{\rm CModel}$.\\
\cname{sigma\_delta\_i} & mag & Uncertainty of $\Delta i$: $\sigma_{\Delta i}=\sqrt{\sigma(i_{\rm PSF})^2+\sigma(i_{\rm CModel})^2}$.\\
\cname{S\_delta\_i} & \nodata & Extendedness significance $S_{\Delta i}=\Delta i/\sigma_{\Delta i}$.\\ \hline
\multicolumn{3}{l}{\textit{Derived radio quantities}}\\ \hline
\cname{alpha\_FV} & \nodata & Spectral index from FIRST--VLASS (1.4--3\,GHz).\\
\cname{alpha\_LV} & \nodata & Spectral index from LoTSS--VLASS (0.15--3\,GHz).\\
\cname{alpha\_LFV} & \nodata & Spectral index from LoTSS--FIRST--VLASS (log--log fit).\\
\cname{alpha\_best} & \nodata & Best-available spectral index (priority: LFV$>$FV$>$LV$>$0.7).\\
\cname{L\_3GHz\_fit\_best} & W\,Hz$^{-1}$ & 3\,GHz luminosity computed with $\alpha_{\rm best}$ using \cname{z\_best}.\\
\cname{L\_3GHz\_fit\_best\_err} & W\,Hz$^{-1}$ & Uncertainty of \cname{L\_3GHz\_fit\_best} propagated using \cname{z\_err\_best}.\\ 
\cname{logR\_i} & dex & \makecell[l]{Radio loudness defined as $\log R_i \equiv \log_{10}(S_{\rm 3\,GHz}/f_i)$.$S_{\rm 3\,GHz}$ is taken from \cname{Total\_flux}.\\
$f_i$ is derived from the Galactic-extinction-corrected HSC $i$-band cModel mag.} 
\enddata
\tablecomments{The full catalog contains additional radio- and optical-measurement columns from the parent surveys. We recommend constructing subsamples with simple radius cuts on \cname{first\_match\_dist\_arcsec} and \cname{lotss\_match\_dist\_arcsec} (Section~\ref{subsec:catalog_subsamples}).}
\end{deluxetable*}

\subsection{Catalog structure and subsamples}\label{subsec:catalog_subsamples}

The catalog is VLASS-centric: each row corresponds to a VLASS component (the primary identifier), and we attach the nearest counterparts in other surveys.
For each VLASS component, we report the nearest HSC counterpart, its separation (\texttt{hsc\_match\_dist\_arcsec}), and the associated HSC properties (with the \texttt{hsc\_} prefix), including the Mizuki photo-$z$ and spec-$z$ information. 
We also provide the nearest-neighbor separations to FIRST and LoTSS (\texttt{first\_match\_dist\_arcsec} and \texttt{lotss\_match\_dist\_arcsec}), together with the corresponding FIRST/LoTSS source properties (with prefixes \texttt{first\_} and \texttt{lotss\_}).
If no counterpart is found within the adopted matching radius, the corresponding match distance and counterpart properties are left blank.
The adopted separation radii and the resulting parent and matched subsamples are summarized in Figure~\ref{flow}; we match VLASS to HSC/FIRST/LoTSS within $1\farcs0/2\farcs5/2\farcs5$ (see Section~3.3). 
We examine the distribution of match separations in Section~5.1.2, where we present and discuss the histogram of the counterpart distances. 

The subsamples defined in Section~3 (blue-outlined boxes in Figure~\ref{flow}) can be reproduced from the released table by applying the following simple cuts. 
Specifically, the radio multi-frequency matched subsamples are obtained via radius cuts:
\begin{itemize}
  \item Clean VLASS--HSC--FIRST: 
  
  \texttt{first\_match\_dist\_arcsec}$\le 2.5$ (18,444 sources).
  
   \item Clean VLASS--HSC--LoTSS: 
  
  \texttt{lotss\_match\_dist\_arcsec}$\le 2.5$ (16,167 sources).
  
  \item Clean VLASS--HSC--FIRST--LoTSS: 
  
  \texttt{first\_match\_dist\_arcsec}$\le 2.5$ and 
   
   \texttt{lotss\_match\_dist\_arcsec}$\le 2.5$ (14,206 sources).
\end{itemize}
Subsamples ``with reliable photo-$z$'' can be reproduced by applying photo-$z$ quality cuts ($\chi^2_\nu<3$ and $\sigma_z<0.1(1+z)$): 
\begin{itemize}
  \item Clean VLASS--HSC with reliable photo-$z$:
  
  \texttt{hsc\_reduced\_chisq} $<3$ and 
  
  \texttt{hsc\_photoz\_std\_best} $<0.1(1+$\texttt{hsc\_photoz\_best}$)$ (15,897 sources).

  \item Clean VLASS--HSC--FIRST with reliable photo-$z$:
  
  \texttt{first\_match\_dist\_arcsec}$\le 2.5$ and 
   
  \texttt{hsc\_reduced\_chisq} $<3$ and 
  
  \texttt{hsc\_photoz\_std\_best} $<0.1(1+$\texttt{hsc\_photoz\_best}$)$ (12,649 sources).

  \item Clean VLASS--HSC--LoTSS with reliable photo-$z$:
  
  \texttt{lotss\_match\_dist\_arcsec}$\le 2.5$ and
  
  \texttt{hsc\_reduced\_chisq} $<3$ and 
  
  \texttt{hsc\_photoz\_std\_best} $<0.1(1+$\texttt{hsc\_photoz\_best}$)$ (11,077 sources).

  \item Clean VLASS--HSC--FIRST--LoTSS with reliable photo-$z$:
  
  \texttt{first\_match\_dist\_arcsec}$\le 2.5$ and
  
  \texttt{lotss\_match\_dist\_arcsec}$\le 2.5$ and
  
  \texttt{hsc\_reduced\_chisq} $<3$ and 
  
  \texttt{hsc\_photoz\_std\_best} $<0.1(1+$\texttt{hsc\_photoz\_best}$)$ (9,627 sources).
\end{itemize}

\subsection{Optical photometry, Mizuki photo-$z$, and spec-$z$}\label{subsec:catalog_photoz}

For HSC counterparts, we provide multi-band forced photometry including \texttt{cmodel} magnitudes and PSF magnitudes (and their uncertainties) in $grizy$.
We also include Galactic extinction coefficients (\texttt{hsc\_a\_\{g,r,i,z,y\}}) and photometric-redshift and stellar-mass estimates from the Mizuki catalog \citep{Tanaka18}.
The primary redshift quantities used in this work are \texttt{hsc\_photoz\_best} and its uncertainty proxy \texttt{hsc\_photoz\allowbreak\_\allowbreak std\_best}.
We also provide stellar-mass estimates based on the spectral energy distribution (SED) fitting to the HSC multi-band photometries \texttt{hsc\_stellar\_mass} and the associated uncertainty ranges (\texttt{hsc\_stellar\_mass\_err68\allowbreak\_\allowbreak min/max}).
When available, we additionally provide spectroscopic redshift information from the HSC-SSP catalog, including \texttt{hsc\_specz} and its quoted uncertainty \texttt{hsc\_specz\_err}.

\begin{figure}
  \centering
  \includegraphics[width=1.0\linewidth]{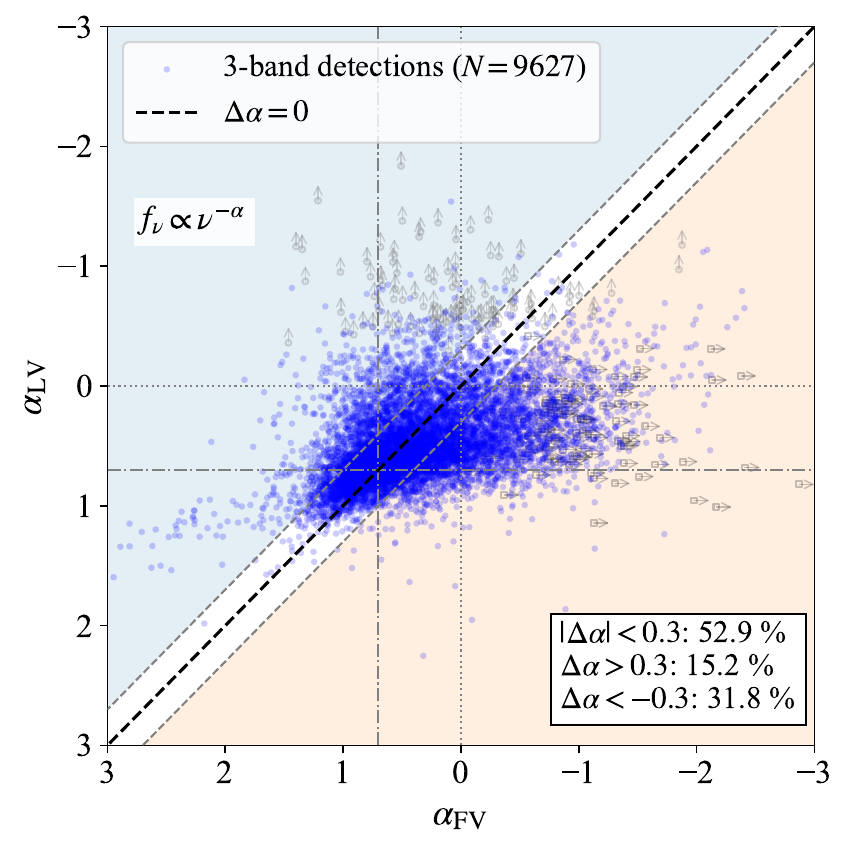}
\caption{
Radio color--color diagram of spectral indices $\alpha_{\rm LV}$ and $\alpha_{\rm FV}$.
Blue points show sources detected in all three radio bands (LoTSS, FIRST, and VLASS), while gray symbols indicate sources detected in only two bands, for which upper limits are shown for the non-detected band. 
Upward arrows represent upper limits on $\alpha_{\rm LV}$ for LoTSS non-detections, and rightward arrows represent upper limits on $\alpha_{\rm FV}$ for FIRST non-detections, assuming representative flux-density limits of 0.35 mJy (LoTSS) and 0.75 mJy (FIRST). 
The dashed line indicates $\alpha_{\rm LV}=\alpha_{\rm FV}$, corresponding to a single power-law spectrum. 
The dotted lines mark $\alpha=0$ and $\alpha=0.7$ for reference. 
The blue- and orange-shaded regions highlight sources with significantly curved radio spectra, defined as $\Delta\alpha<-0.3$ (convex) and $\Delta\alpha>0.3$ (concave), respectively. 
The fractions of sources consistent with power-law ($|\Delta\alpha|<0.3$), convex ($\Delta\alpha>0.3$), and concave ($\Delta\alpha<-0.3$) spectra are computed using only the three-band detections and are shown in the lower-right corner. 
For clarity, only a random subset of 100 sources is shown for each class of upper limits, corresponding to $\sim3\%$ (LoTSS) and $\sim7\%$ (FIRST) of the full sample. 
}
  \label{fig:alpha_alpha}
\end{figure}

\subsection{Derived quantities: spectral indices}\label{subsec:catalog_alpha}

We define radio spectral indices using $S_\nu \propto \nu^{-\alpha}$.
Using integrated flux densities at 144\,MHz (LoTSS), 1.4\,GHz (FIRST), and 3\,GHz (VLASS), we compute:
(i) $\alpha_{\rm FV}$ from FIRST--VLASS (1.4--3\,GHz),
(ii) $\alpha_{\rm LV}$ from LoTSS--VLASS (0.15--3\,GHz), and
(iii) $\alpha_{\rm LFV}$ from a linear fit in $\log S$--$\log\nu$ using LoTSS, FIRST, and VLASS when all three fluxes are available.
As a simple spectral-curvature indicator, we also provide
$\Delta\alpha \equiv \alpha_{\rm FV}-\alpha_{\rm LV}$ (stored as \texttt{delta\_alpha}). 

We define a best-available spectral index $\alpha_{\rm best}$ (stored as \texttt{alpha\_best}) by prioritizing the broadest frequency coverage: 
\begin{equation}
\alpha_{\rm best} =
\left\{
\begin{array}{ll}
\alpha_{\rm LFV}, & \mbox{if $\alpha_{\rm LFV}$ is available},\\
\alpha_{\rm FV},  & \mbox{else if $\alpha_{\rm FV}$ is available},\\
\alpha_{\rm LV},  & \mbox{else if $\alpha_{\rm LV}$ is available},\\
0.7,              & \mbox{otherwise}.\\
\end{array}
\right.
\end{equation}

 \begin{figure*}
  \centering
  \includegraphics[width=\textwidth]{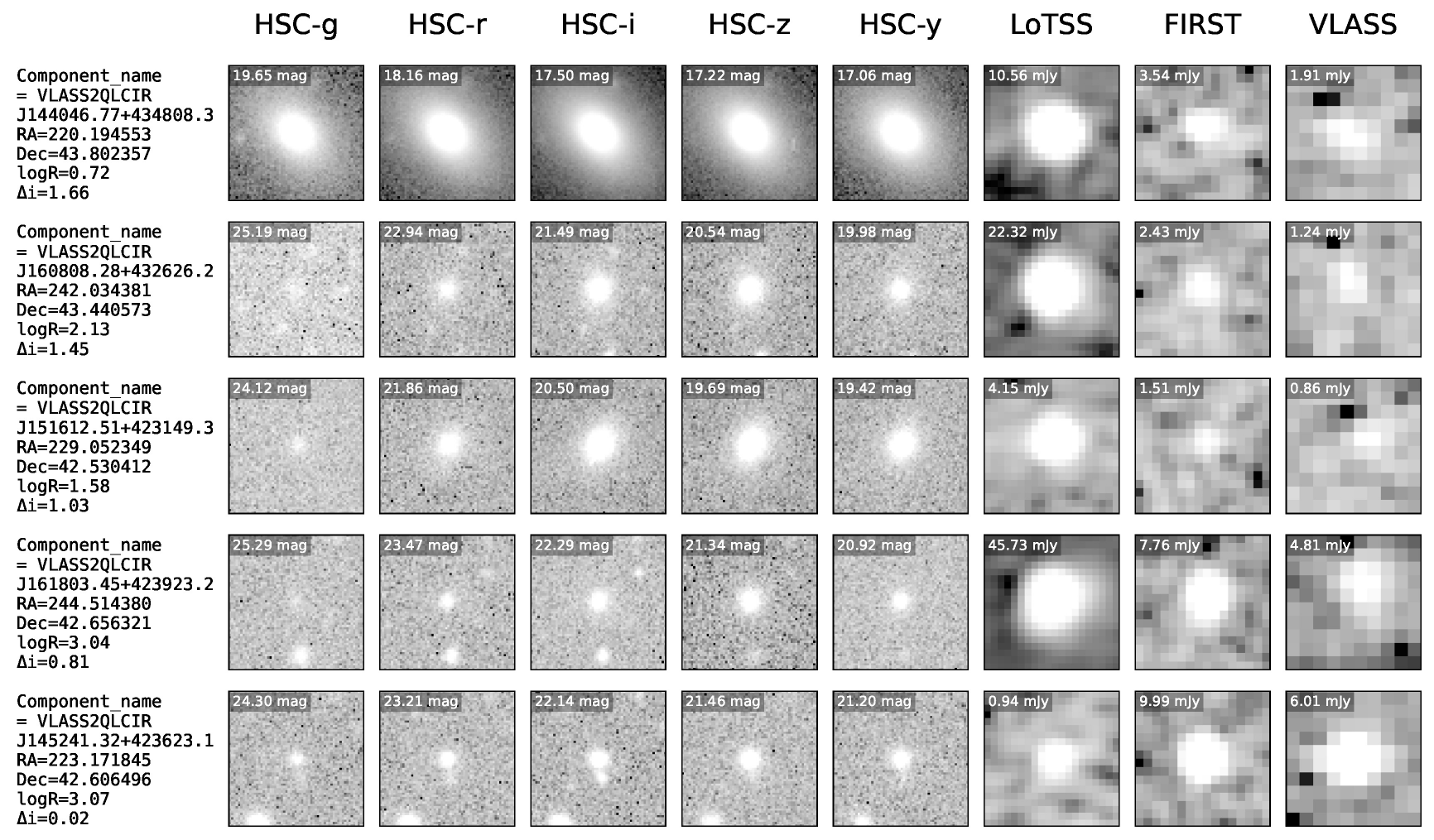}
\caption{
Multi-wavelength cutout montage for five representative sources from the Clean VLASS--HSC--FIRST--LoTSS sample.
Each row corresponds to one source, and the columns show the HSC $g$, $r$, $i$, $z$, and $y$-band images, followed by radio cutouts from LoTSS (144\,MHz), FIRST (1.4\,GHz), and VLASS (3\,GHz).
The HSC and VLASS panels are $10^{\prime\prime}\times10^{\prime\prime}$ cutouts, while the LoTSS and FIRST panels are $30^{\prime\prime}\times30^{\prime\prime}$ cutouts to better visualize the radio morphology in the lower-resolution radio images.  All cutouts are centered on the HSC counterpart position. 
For the HSC bands, the overlaid label indicates the AB magnitude, while for the radio bands it indicates the integrated flux density in mJy.
The left-hand annotation lists the VLASS component name, sky coordinates, radio loudness (log$R$; defined in Section~\ref{subsec:catalog_logRi}), and optical extendedness  ($\Delta i$; defined in Section~\ref{subsec:catalog_deltai}) for each source.
}
\label{fig:cutout_montage}
\end{figure*}

To investigate the spectral shapes of radio sources, we construct a radio color--color diagram using the spectral indices $\alpha_{\rm LV}$ and $\alpha_{\rm FV}$, as shown in Figure~\ref{fig:alpha_alpha}. 
The majority of the three-band detections lie close to the one-to-one relation ($\alpha_{\rm LV}=\alpha_{\rm FV}$), indicating that their radio spectra are broadly consistent with a single power-law. 
We find that $52.9\%$ of the sources satisfy $|\Delta\alpha|<0.3$. 
In contrast, $15.2\%$ and $31.8\%$ of the sources show $\Delta\alpha>0.3$ and $\Delta\alpha<-0.3$, respectively, corresponding to convex and concave spectral shapes.
Sources with upper limits exhibit a broader distribution. They are not included in the quantitative classification due to the lack of well-constrained spectral indices, but they qualitatively support the presence of a diverse range of spectral shapes in the full sample.

\subsection{Derived quantities: 3\,GHz radio luminosities}\label{subsec:catalog_lum}

We compute the rest-frame monochromatic radio luminosity at 3\,GHz, $L_{\rm 3\,GHz}$ (W\,Hz$^{-1}$), using the observed VLASS integrated flux density $S_{\rm 3\,GHz}$ and a $K$-correction based on the spectral index $\alpha$:
\begin{equation}
L_{\rm 3\,GHz} = 4\pi D_L(z)^2\, S_{\rm 3\,GHz}\, (1+z)^{\alpha-1},
\end{equation}
where $D_L(z)$ is the luminosity distance.
For the luminosity calculation, we adopt a best-available redshift, \texttt{z\_best}, which prioritizes spectroscopic redshifts when available and otherwise falls back to photometric redshifts.
Specifically, we set \texttt{z\_best}=\texttt{hsc\_specz} (and \texttt{z\_err\_best}=\texttt{hsc\_specz\_err}) when a spectroscopic redshift is present, and otherwise use \texttt{z\_best}=\texttt{hsc\_photoz\_best} (with \texttt{z\_err\_best}=\texttt{hsc\_photoz\_std\_best}).

In the released table, we provide luminosities computed with a fixed $\alpha=0.7$ (\texttt{L\_3GHz\_WHz\_with\_a\_0p7}) as well as values computed using $\alpha_{\rm FV}$, $\alpha_{\rm LV}$, $\alpha_{\rm LFV}$, and $\alpha_{\rm best}$ (columns \texttt{L\_3GHz\_fit\_FV/LV/LFV/best}).
We propagate uncertainties from the VLASS flux error and \texttt{z\_err\_best} to provide corresponding luminosity uncertainties.\footnote{
Stellar-mass estimates included in the table are taken from the Mizuki catalog and are based on the Mizuki photometric-redshift solution.
They are not recomputed using spectroscopic redshifts when available; therefore, for sources where photo-$z$ and spec-$z$ are inconsistent, the stellar-mass estimate may be biased.
We recommend applying the reliable photo-$z$ selection (Section~3; Figure~\ref{flow}) and, when using spec-$z$ sources, additionally requiring consistency between photo-$z$ and spec-$z$.
}

\subsection{Derived quantities: optical extendedness}\label{subsec:catalog_deltai}

To illustrate how far HSC imaging can identify spatially extended counterparts at high redshift,
we use the $i$-band morphology diagnostic
$\Delta i \equiv i_{\rm PSF}-i_{\rm CModel}$, where 
$i_{\rm PSF}$ and $i_{\rm CModel}$ denote the HSC $i$-band PSF and \texttt{cModel} magnitudes, respectively.
By construction, $\Delta i>0$ indicates that the source is more extended than the PSF and hence
is spatially resolved by HSC in the $i$ band.

Within the Clean VLASS--HSC parent sample (restricted to objects with valid
$i_{\rm PSF}$ and $i_{\rm CModel}$ measurements), we quantify the significance of the extendedness as
\begin{equation}
S_{\Delta i}=\frac{\Delta i}{\sigma_{\Delta i}},
\end{equation}
where the uncertainty is propagated from the reported magnitude errors as
\begin{equation}
\sigma_{\Delta i}=\sqrt{\sigma(i_{\rm PSF})^2+\sigma(i_{\rm CModel})^2}.
\end{equation}
We provide the derived quantities $\Delta i$, $\sigma_{\Delta i}$, and $S_{\Delta i}$ in the columns
\texttt{delta\_i}, \texttt{sigma\_delta\_i}, and \texttt{S\_delta\_i}.

\subsection{Derived quantities: radio loudness}\label{subsec:catalog_logRi}

To quantify the relative strength of radio emission with respect to the optical host,
we define the $i$-band radio loudness as
\begin{equation}
\log R_i \equiv \log_{10}\!\left(\frac{S_{\rm 3\,GHz}}{f_i}\right),
\end{equation}
where $S_{\rm 3\,GHz}$ is the VLASS 3\,GHz integrated (total) flux density
(\texttt{Total\_flux}; in Jy) and $f_i$ is the $i$-band flux density (in Jy)
derived from the Galactic-extinction-corrected HSC \texttt{cModel} magnitude,
$i_{\rm CModel,corr}=i_{\rm CModel}-A_i$, assuming the AB system:
\begin{equation}
f_i = 3631~{\rm Jy}\times 10^{-0.4\,i_{\rm CModel,corr}}.
\end{equation}
We provide $\log R_i$ in the catalog column \texttt{logR\_i}.

\subsection{Sample example}

Figure~\ref{fig:cutout_montage} shows a set of representative sources drawn from the Clean VLASS--HSC--FIRST--LoTSS sample, illustrating the typical optical morphologies and radio morphologies across a range of radio loudness and optical extendedness. 
These examples demonstrate that a non-negligible fraction of our radio-selected AGN candidates reside in clearly extended host galaxies in the HSC images while exhibiting diverse radio morphologies in LoTSS, FIRST, and VLASS.


\section{Catalog Validation and quality assessment}

\subsection{Photometric redshift quality and validation with spectroscopic redshifts}
\label{subsec:photoz_validation}

\begin{figure}
  \centering
  \includegraphics[width=1.0\linewidth]{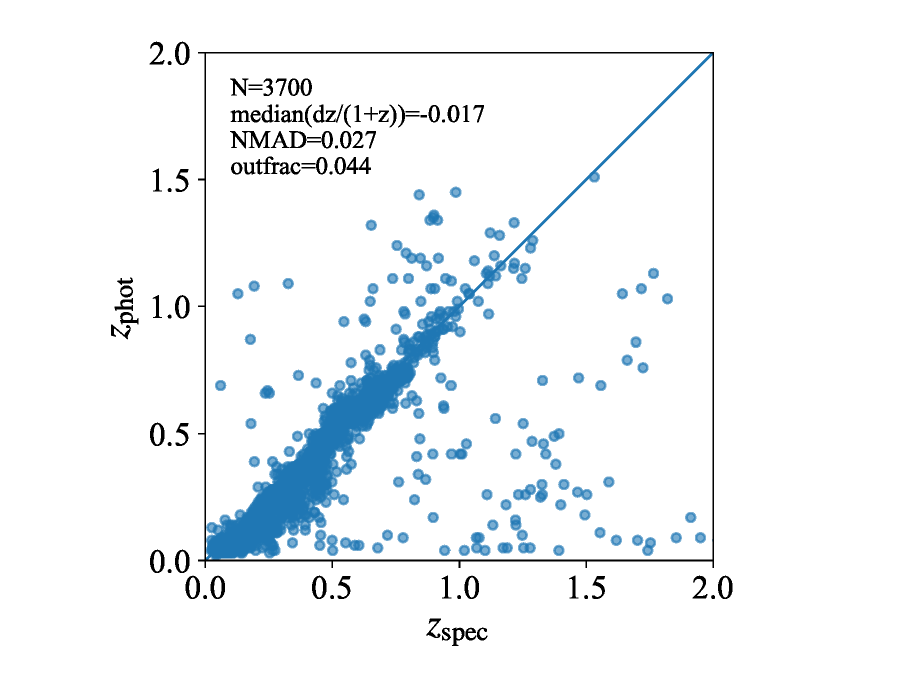}
  \caption{
  Comparison of spectroscopic and photometric redshifts for sources in our catalog.
  The solid line indicates $z_{\rm phot}=z_{\rm spec}$.
  The inset text reports the number of matched sources and the summary statistics of
  $\Delta z/(1+z_{\rm spec})$, including the median bias, NMAD scatter, and the outlier fraction
  (defined by $|\Delta z|/(1+z_{\rm spec})>0.15$).
  }
  \label{fig:specz_vs_photoz}
\end{figure}

\begin{figure}
  \centering
  \includegraphics[width=1\linewidth]{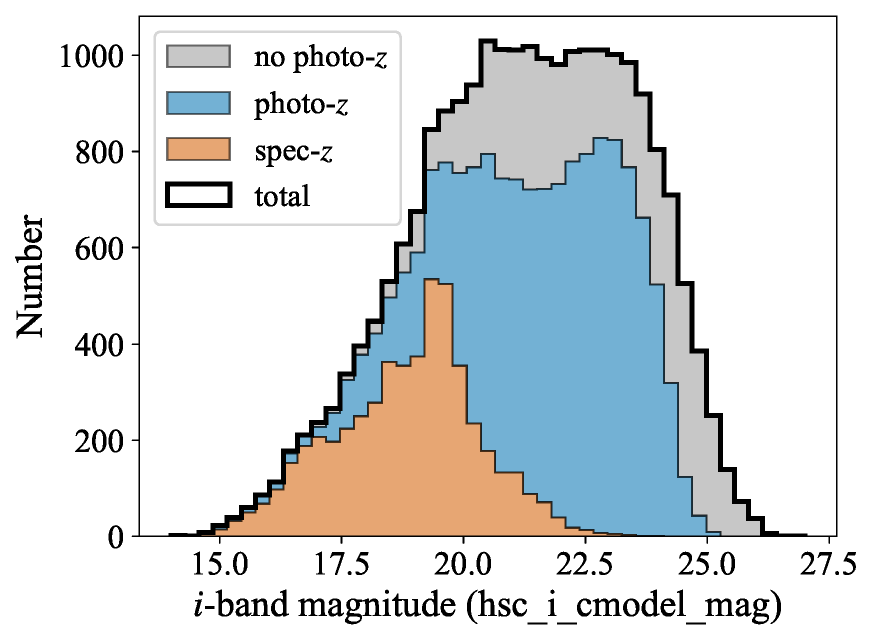}
\caption{
Stacked histogram of HSC $i$-band cModel magnitudes for the Clean VLASS--HSC sample (black outline; total base sample with finite $i$ magnitude). 
Sources are partitioned into three exclusive redshift-availability categories with priority, 
spec-$z >$ photo-$z >$ no photo-$z$:
(i) objects with a secure spectroscopic redshift (orange),
(ii) objects without spec-$z$ but with a reliable photometric redshift satisfying
$\chi^2_\nu < 3$ and $\sigma_z < 0.1(1+z)$ (blue), and (iii) all remaining objects (non-reliable photo-$z$) (gray).
}
  \label{fig:iband_hist}
\end{figure}

To validate the photometric-redshift estimates for the galaxies in our VLASS--HSC catalog,
we compare the Mizuki photo-$z$ values  $z_{\rm phot}$ to spectroscopic redshifts $z_{\rm spec}$. 
Figure~\ref{fig:specz_vs_photoz} shows the comparison between $z_{\rm spec}$ and $z_{\rm phot}$. 
The normalized redshift difference $\Delta z/(1+z_{\rm spec}) \equiv (z_{\rm phot}-z_{\rm spec})/(1+z_{\rm spec})$ has a median bias of $-0.017$, a normalized median absolute deviation (NMAD) scatter of $0.027$, and an outlier fraction of $0.044$ (using the conventional threshold $|\Delta z|/(1+z_{\rm spec})>0.15$).
These statistics indicate that the photo-$z$ estimates are broadly reliable for our purposes,
while a small population of catastrophic outliers remains.

The asymmetry in the outlier distribution, with more sources showing $z_{\rm phot}<z_{\rm spec}$, is mainly seen for sources at $z_{\rm spec}\gtrsim1.3$. At these redshifts, the 4000\,\AA\ break is redshifted to $\gtrsim9200$\,\AA, near the reddest HSC bands or beyond the HSC wavelength coverage. 
As a result, the HSC $grizy$ photometry alone provides only weak constraints on the break, leading to degeneracies in template fitting and preferentially lower-redshift solutions. 
We note that Mizuki is primarily optimized for galaxy SEDs, and the photometric redshifts may therefore be less reliable for AGN-dominated sources such as broad-line quasars.


Figure~\ref{fig:iband_hist} shows the $i$-band magnitude distributions for the Clean VLASS--HSC sample. 
At the bright end (and correspondingly at lower redshift), the sample is largely determined by objects with secure spectroscopic redshifts, indicating that redshift information in this regime is dominated by spec-$z$ coverage. 
In contrast, toward higher redshift the population with reliable photometric redshifts becomes increasingly dominant, implying that the redshift census at high-$z$ is primarily set by photo-$z$ availability rather than spectroscopy.

At the faint end, the distribution is dominated by sources without a reliable photo-$z$. 
These sources are likely associated with Lyman-break--type (dropout) populations for which standard photo-$z$ fits are prone to fail or have large uncertainties. 
A detailed investigation of this faint, dropout population will be presented in \citet{Kong2026}.

\begin{figure}
  \centering
  \includegraphics[width=1\linewidth]{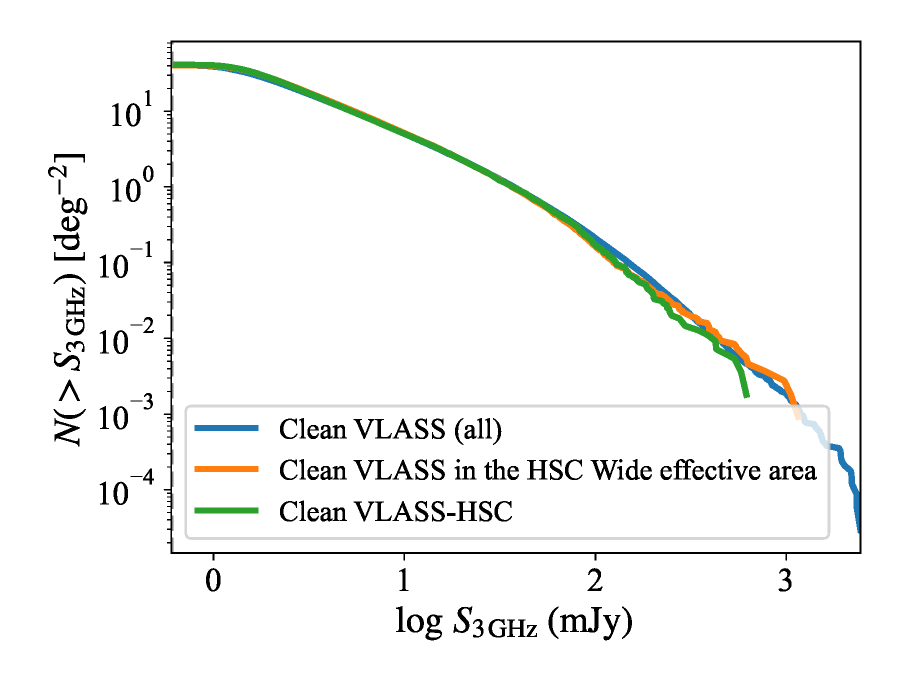}
\caption{
Cumulative source number densities as a function of 3\,GHz flux density for the Clean VLASS catalog (blue), Clean VLASS sources within the HSC Wide effective area (orange), and the Clean VLASS--HSC sample (green). 
The vertical axis shows the cumulative surface density $N(>S)$ in units of deg$^{-2}$. 
}
  \label{fig:radio_flux_cumulative_comparison}
\end{figure}

\begin{figure*}
  \centering
  \includegraphics[width=\textwidth]{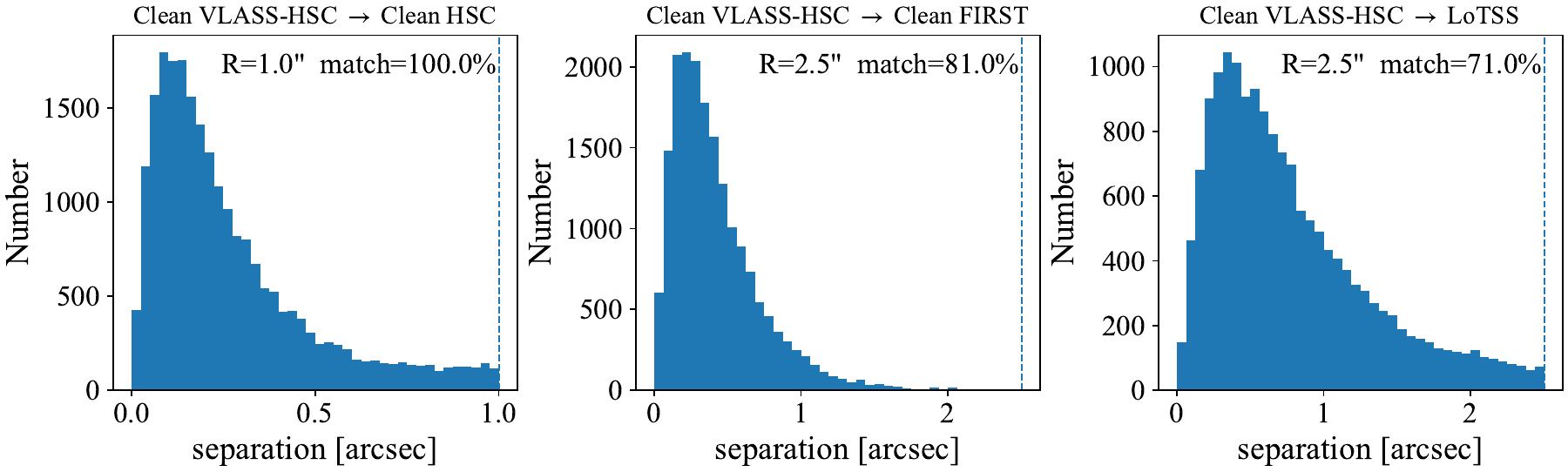}
  \caption{
Histograms of nearest-neighbor positional-match separations $d$ for the Clean VLASS--HSC parent sample to Clean HSC (left; matched within $1\farcs0$), Clean FIRST (middle; matched within $2\farcs5$), and LoTSS (right; matched within $2\farcs5$). 
Only sources within the adopted matching radii are shown in each panel. 
The dashed vertical lines indicate the adopted matching radii.
}
  \label{fig:sep_hist_3panel}
\end{figure*}

\begin{figure}
  \centering
  \includegraphics[width=1\linewidth]{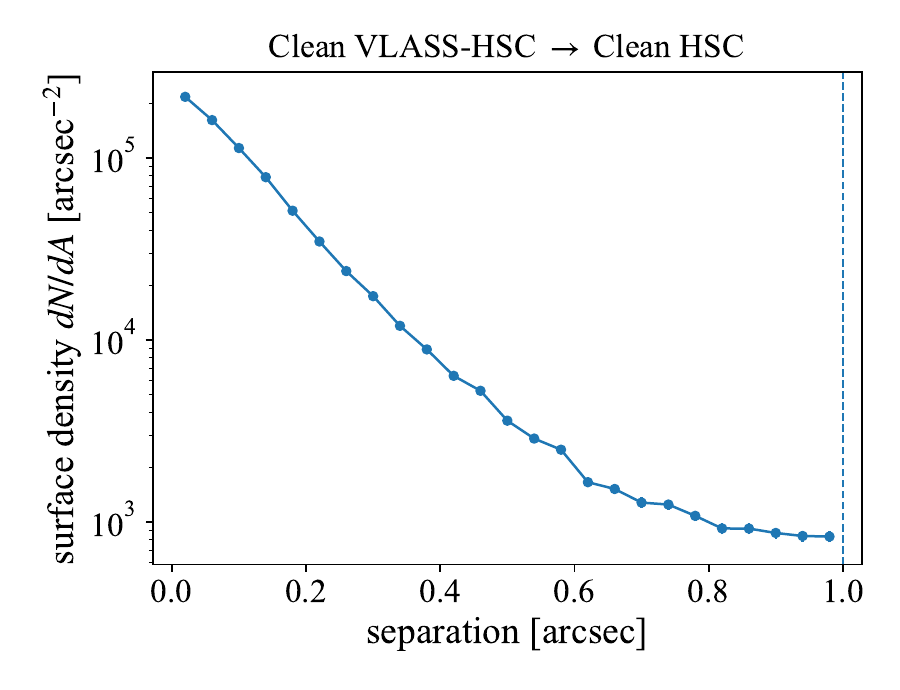}
\caption{
Radial surface density of VLASS--HSC matches.
The surface density of HSC counterparts around VLASS positions, $dN/dA$, is shown as a function of angular separation.
The vertical dashed line marks the adopted matching radius of $1.0''$.
}
  \label{fig:vlass_hsc_radial_density}
\end{figure}

\subsection{Check for radio-flux bias introduced by the optical matching} 
To examine whether the optical cross-matching introduces a bias in radio flux, we compare the radio flux distributions of the full Clean VLASS sample, the subset located within the HSC--SSP Wide effective area, and the Clean VLASS--HSC sample in Figure~\ref{fig:radio_flux_cumulative_comparison}.
The three cumulative distributions are very similar over most of the flux range.
A Kolmogorov--Smirnov (KS) test also shows only small differences, with KS statistics in the range KS$=0.014$--$0.048$.
In particular, the sample within the HSC Wide effective area and the Clean VLASS--HSC sample are nearly identical (KS$=0.014$), indicating that the optical matching does not introduce a strong bias in radio flux.

\subsection{Positional accuracies and match-separation distributions}
Figure~\ref{fig:sep_hist_3panel} shows the distributions of angular separations for our positional cross-matches.
By construction, all sources in the Clean VLASS--HSC catalog have an HSC counterpart within $1\farcs0$.
The VLASS--HSC separations are tightly peaked at small offsets, with a median of $0\farcs199$ and 95\% of the separations within $0\farcs771$ (Table~\ref{tab:sep_stats}).
For FIRST, 18,444 sources (81.0\% of the Clean VLASS--HSC sample) have counterparts within $2\farcs5$, with a median separation of $0\farcs345$.
For LoTSS, 16,167 sources (71.0\% of the Clean VLASS--HSC sample) are matched within the same radius, with a median separation of $0\farcs637$.
These separation distributions indicate that our cross-matching procedure yields reliable positional associations across the three surveys.

For the VLASS--HSC matches, however, Figure~\ref{fig:sep_hist_3panel} also suggests that the separation distribution becomes much flatter toward larger radii.
To examine this behavior more directly, Figure~\ref{fig:vlass_hsc_radial_density} shows the radial surface density of HSC sources around VLASS positions, $dN/dA$, as a function of angular separation.
The surface density shows a pronounced central excess at small separations, whereas it approaches an approximately constant level at larger separations.
This outer, nearly constant component is consistent with the contribution from random associations with unrelated HSC sources.
The adopted matching radius of $1.0''$ therefore includes the bulk of genuine counterparts while limiting contamination from chance alignments.
This interpretation is also consistent with the empirical contamination estimate presented in Appendix~A, which gives $f_{\rm contam}(1.0'') \lesssim 0.1$.

\begin{deluxetable*}{lcccccccc}
\tablecaption{Summary of match separations\label{tab:sep_stats}}
\tabletypesize{\scriptsize}
\tablehead{
\colhead{Match} &
\colhead{$R_{\rm match}$} &
\colhead{$N_{\rm total}$} &
\colhead{$N_{\rm matched}$} &
\colhead{Frac.} &
\colhead{Min} &
\colhead{Median} &
\colhead{P95} &
\colhead{Max}
}
\startdata
Clean VLASS--HSC $\rightarrow$ Clean HSC         & $1\farcs0$ & 22,773 & 22,773 & 1.0000 & $0\farcs001175$ & $0\farcs198871$ & $0\farcs770819$ & $0\farcs999627$ \\
Clean VLASS--HSC $\rightarrow$ Clean FIRST     & $2\farcs5$ & 22,773 & 18,444 & 0.8099 & $0\farcs001152$ & $0\farcs344855$ & $1\farcs014607$ & $2\farcs481731$ \\
Clean VLASS--HSC $\rightarrow$ LoTSS         & $2\farcs5$ & 22,773 &  16,167 & 0.7099 & $0\farcs005072$ & $0\farcs636960$ & $1\farcs942074$ & $2\farcs497826$ \\
\enddata
\tablecomments{
Using the Clean VLASS--HSC parent catalog, we compute the nearest-neighbor angular separation $d$ to the closest counterpart in each external catalog: Clean HSC, Clean FIRST, and LoTSS.
We adopt a match radius $R_{\rm match}$ and define matched sources as those with $d \le R_{\rm match}$.
The columns Min--Max and the percentile ($P95$) are computed from the matched-only separation distribution (i.e., using only sources with $d \le R_{\rm match}$).
}
\end{deluxetable*}

\begin{figure*}
  \centering
  \includegraphics[width=2.0\columnwidth]{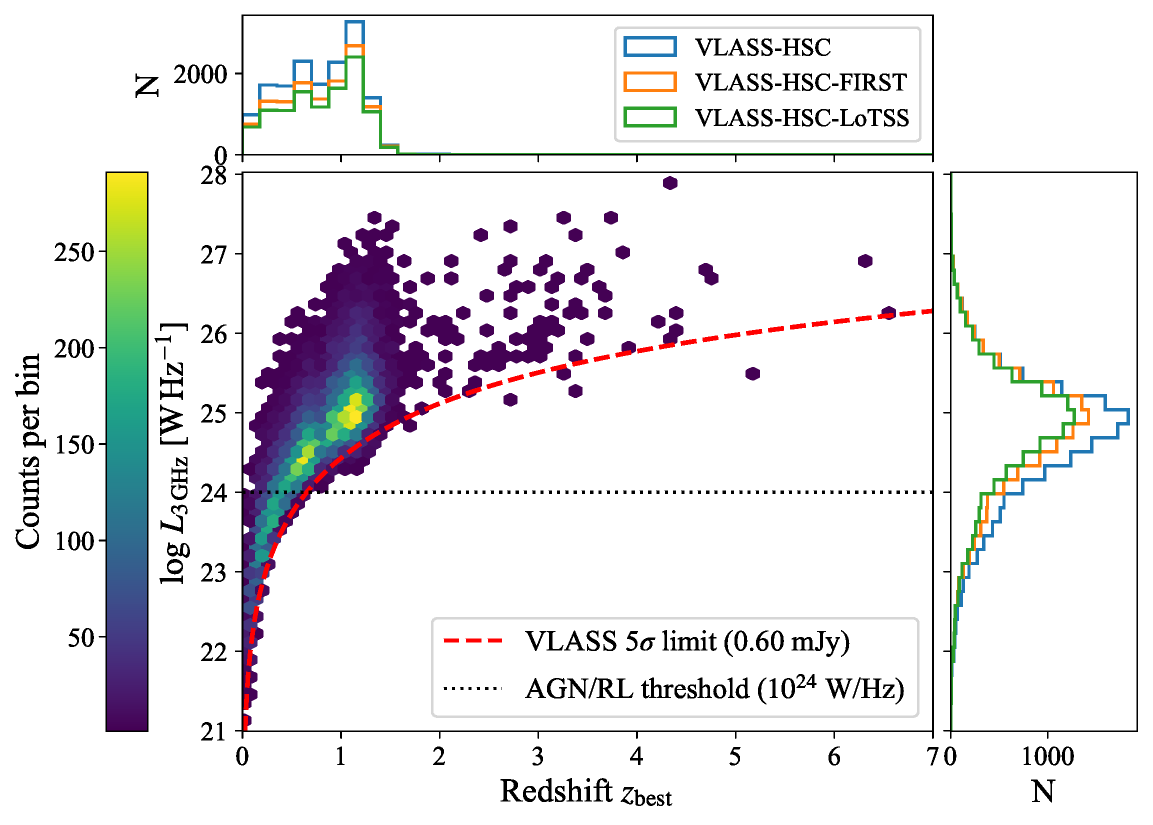}
\caption{
Radio luminosity at 3~GHz ($L_{\rm 3\,GHz}$) versus redshift for the Clean VLASS--HSC with reliable photo-$z$ catalog.
The central panel shows the distribution in the $z_{\rm best}$--$\log (L_{\rm 3\,GHz}/{\rm W\,Hz^{-1}})$ plane,
where $z_{\rm best}$ is derived from \texttt{z\_best}, 
and $L_{\rm 3\,GHz}$ is derived from \texttt{L\_3GHz\_fit\_best}.
The density is displayed as a hexbin map (color indicates counts per bin).
The top and right panels show the marginal distributions of $z_{\rm best}$ and $\log L_{\rm 3\,GHz}$
for three subsamples: Clean VLASS--HSC with reliable photo-$z$ (blue), Clean VLASS--HSC--FIRST with reliable photo-$z$ (orange), and Clean VLASS--HSC--LoTSS with reliable photo-$z$ (green). 
The red dashed curve indicates the VLASS $5\sigma$ flux-density limit (0.60\,mJy at 3\,GHz) converted to a luminosity limit as a function of redshift assuming a spectral index $\alpha=0.7$ ($S_\nu \propto \nu^{-\alpha}$).
The black dotted line marks the commonly adopted radio-loud/AGN threshold,
$L_{\rm 3\,GHz}=10^{24}\,{\rm W\,Hz^{-1}}$.
}
  \label{fig:z_logL3}
\end{figure*}

\section{Catalog Properties}

\subsection{Redshift and Radio-Luminosity Distributions}

\begin{figure*}
  \centering
  \includegraphics[width=\linewidth]{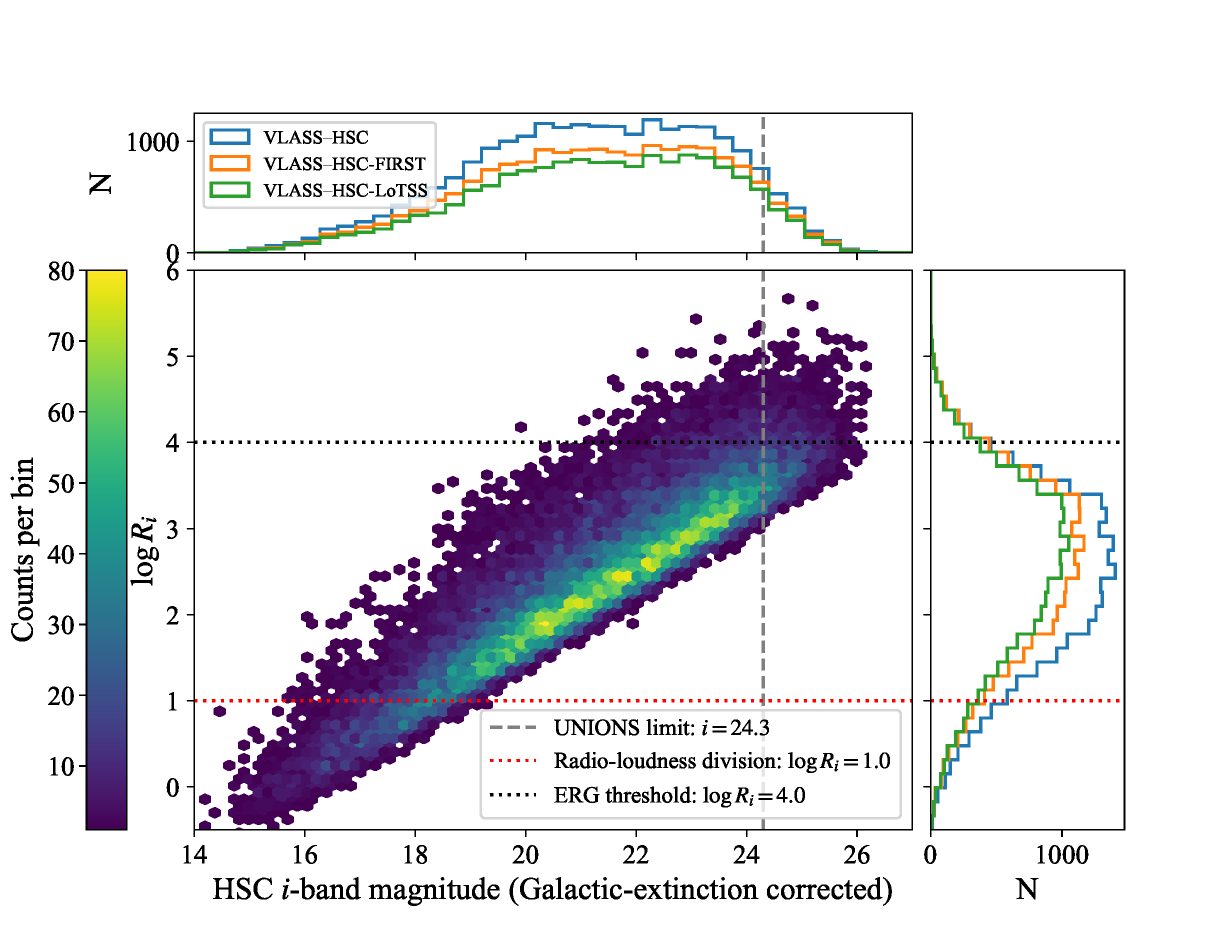}
  \caption{
  Radio-to-optical ratio as a function of optical brightness for our radio-selected samples.
  The central panel shows the distribution of the radio-to-optical ratio, $\log R_i \equiv \log (S_{\rm 3\,GHz}/f_i)$,
  versus the Galactic-extinction-corrected HSC $i$-band cModel magnitude for the Clean VLASS--HSC sample after requiring ${\rm S/N}(i)\ge 5$.
  The density is displayed using a hexbin representation, with the color scale indicating
  the number of sources per bin.
  The top and right panels show the marginal distributions of $i$ and $\log R_i$, respectively,
  for the Clean VLASS--HSC sample (blue), Clean VLASS--HSC--FIRST (orange) and Clean VLASS--HSC--LoTSS (green).
  The vertical dashed line marks the representative UNIONS $i$-band limiting magnitude
  ($i=24.3$), while the horizontal black dotted line indicates an ``extremely radio-loud'' threshold
  at $\log R_i=4$ (see text). 
  We also show the conventional radio-loudness division at $\log R_i=1$ as a red dotted horizontal line. 
  \label{fig:i_logR_joint}}
\end{figure*}

Figure~\ref{fig:z_logL3} presents the distribution of rest-frame 3 GHz radio luminosity as a function of best-available redshift ($z_{\rm best}$) for our radio-selected samples. 
The three radio-selected samples---the Clean VLASS--HSC with reliable photo-$z$, Clean VLASS--HSC--FIRST with reliable photo-$z$, and Clean VLASS--HSC--LoTSS with reliable photo-$z$ samples---are dominated at $z\lesssim 1.5$ and exhibit broadly similar redshift distributions. 
This behavior is expected given that the HSC photometric redshifts are derived from the five broad bands ($grizy$).
For typical galaxy SEDs, the 4000 \AA~ break can be tracked robustly within the HSC wavelength coverage up to $z\sim1.5$ (i.e., it remains within the $y$ band), which naturally yields a redshift distribution dominated at $z\lesssim1.5$.
At higher redshift the 4000 \AA~ break shifts beyond the $y$ band, and photo-$z$ constraints from $grizy$ alone become less discriminating. 
Table~\ref{tab:z_percentiles} summarizes the median and upper-tail percentiles of the Mizuki photo-$z$ distributions.
The medians are $z_{\rm med}\simeq 0.74$--0.77, while the 90th percentiles are $z_{90}\simeq 1.24$--1.25.
A small high-redshift tail is present, with $z_{99}\simeq 1.99$--2.01 depending on the subsample.
These objects are rare but preferentially luminous, consistent with the strongly flux-limited nature of VLASS at high redshift.
However, we caution that photometric redshifts can be biased to low-$z$ solutions, often missing 
genuine high-$z$ sources.
In particular, a fraction of sources selected as high-redshift dropout radio AGN are assigned relatively low $z_{\rm phot}$ by Mizuki (see Kong et al. submitted.).
This discrepancy can be interpreted as most likely arising from the photo-$z$ prior, which can favor low-redshift solutions when the photometry is ambiguous.

\begin{deluxetable}{lccc}
\tablecaption{Photometric-redshift percentiles (Mizuki)\label{tab:z_percentiles}}
\tabletypesize{\scriptsize}
\tablehead{
\colhead{Sample} & \colhead{$z_{\rm med}$} & \colhead{$z_{90}$} & \colhead{$z_{99}$}
}
\startdata
Clean VLASS--HSC               & 0.74 & 1.24 & 1.99 \\
Clean VLASS--HSC--FIRST        & 0.75 & 1.25 & 2.01 \\
Clean VLASS--HSC--LoTSS        & 0.77 & 1.25 & 2.00 \\
\enddata
\tablecomments{Percentiles are computed for sources with valid Mizuki photo-$z$ measurements in each sample.}
\end{deluxetable}

The central panel of Figure~\ref{fig:z_logL3} further illustrates how the radio-luminosity distribution depends on redshift.
The bulk of the population locate at $\log (L_{\rm 3\,GHz}/\mathrm{W~Hz^{-1}}) \sim 24$--25, with a tail extending to
$\log (L_\mathrm{3\,GHz}/\mathrm{W\,Hz^{-1}}) \gtrsim 26$ (see also the right panel of Figure~\ref{fig:z_logL3}).
The red dashed curve shows the luminosity corresponding to the VLASS $5\sigma$ detection limit as a function of redshift.
As expected, this selection boundary produces a lower envelope that rises rapidly toward high $z$, 
so that intrinsically radio-quiet sources are progressively missed at $z\gtrsim 1$. 
This behavior reflects the flux-limited nature of the sample, i.e., a manifestation of the Malmquist bias. 
Note that the VLASS $5\sigma$ limit shown in the figure assumes a spectral index of $\alpha=0.7$, while the plotted spectral indices are often derived from fits to VLASS/FIRST/LoTSS flux densities; this difference causes some sources to appear below the nominal limit.

Radio emission at $L_{\nu}\gtrsim 10^{24}\,{\rm W\,Hz^{-1}}$ is typically dominated by radio AGN rather than star formation
(e.g., \citealt{MauchSadler2007,BestHeckman2012}; see also \citealt{Smolcic2017A1} for deep 3\,GHz surveys), while radio luminosity alone is not a definitive AGN classifier. 
Adopting $L_{\rm 3\,GHz}=10^{24}\,{\rm W\,Hz^{-1}}$ as a practical divider, 
we find that a large fraction of the Clean VLASS--HSC sample lies above this threshold:
82.1\% sources satisfy $\log (L_{\rm 3\,GHz}/\mathrm{W\,Hz^{-1}})\ge 24$.

\begin{figure*}
  \centering
  \includegraphics[width=\textwidth]{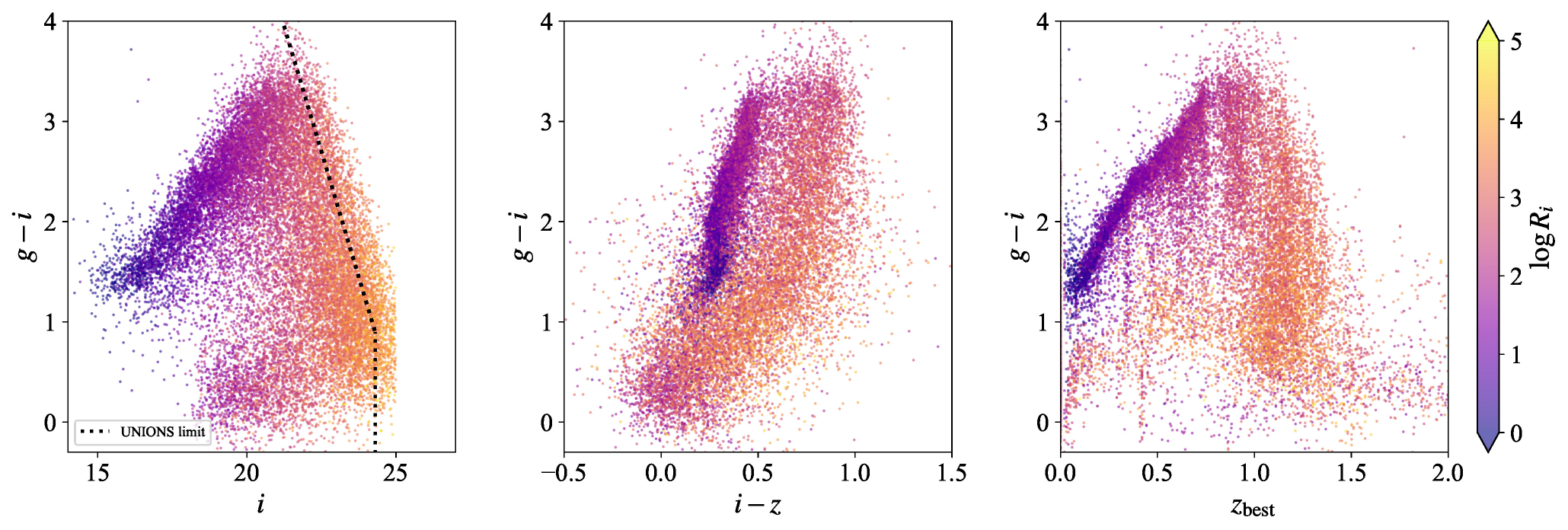}
\caption{
Optical color distributions of the Clean VLASS--HSC sample in the $g-i$ versus $i$ color--magnitude diagram (left), the $g-i$ versus $i-z$ color--color diagram (middle), and the $g-i$ versus $z_{\rm best}$,  color-coded by the radio-to-optical ratio
$\log R_i \equiv \log (S_{\rm 3\,GHz}/f_i)$, 
where $S_{\rm 3\,GHz}$ is the observed 3\,GHz flux density from VLASS and $f_i$ is the $i$-band flux density
derived from the HSC $i$-band magnitude. 
All $g$, $i$, and $z$ magnitudes are corrected for Galactic extinction. 
We require $S/N \ge 5$ in the HSC $g$, $i$, and $z$ bands to ensure robust colors.
The black dashed boundary indicates the approximate UNIONS detection limit, corresponding to representative depths of $i \simeq 24.3$ and $g \simeq 25.2$ ($5\sigma$), shown in the $i$ versus $g-i$ plane. 
  \label{fig:gi_cmd_cc}}
\end{figure*}

\subsection{Radio loudness Distributions}\label{sec:logR_def}

Figure~\ref{fig:i_logR_joint} summarizes the relationship between optical brightness and radio loudness for our radio-selected sources.
To ensure robust optical measurements, we restrict the analysis to sources with ${\rm S/N}(i)\ge 5$ in the Clean VLASS-HSC catalog.

The distribution in Figure \ref{fig:i_logR_joint} shows a clear positive trend between $i$-band magnitude and $\log R_i$ defined in \S\ref{subsec:catalog_logRi}:
as expected for a flux-limited radio sample, optically fainter counterparts tend to have larger radio-to-optical ratios.
Despite this overall trend, the distributions of the Clean VLASS-HSC-FIRST and Clean VLASS-HSC-LoTSS subsamples largely follow the same locus as the parent Clean VLASS--HSC sample, indicating that additional radio cross-matches reduce only the sample size rather than selecting a fundamentally different optical--radio regime. 

Quantitatively, the Clean VLASS--HSC sample with ${\rm S/N}(i)\ge 5$ contains $N=22,681$ sources, of which $N(\log R_i>1)=20,870$ satisfy a conventional ``radio-loud'' criterion based on the radio-to-optical ratio.
Adopting $\log R_i=4$ as a practical threshold for an extremely radio-loud population \citep[ERG;][]{Ichikawa2021WERGS4}, 
we identify $N(\log R_i>4)=950$ sources. 
For comparison with wide-area optical surveys such as UNIONS, 
approximately 50 \% of these extremely radio-loud objects are bright enough to be within the representative UNIONS $i$-band depth ($i\le 24.3$), with $N(\log R_i>4,\ i\le 24.3)=497$.
The remaining $\log R_i>4$ sources are optically fainter than the representative UNIONS limit and are therefore
preferentially accessible via deeper optical imaging such as HSC-SSP, highlighting the value of
HSC-based identifications for assembling large samples of high radio-to-optical ratio systems. 





\subsection{Optical properties}

Figure~\ref{fig:gi_cmd_cc} summarizes the optical properties of the Clean VLASS--HSC sources
using the $g$ -- $i$ versus $i$ color--magnitude diagram (CMD), the $g-i$ versus $i-z$ color--color plane, and the $g-i$ color as a function of redshift, with points color-coded by the radio-to-optical ratio $\log R_i$.
To minimize spurious color outliers, we restrict the analysis to objects in the Clean VLASS-HSC catalog with $S/N\ge 5$ in the HSC $g$, $i$, and $z$ bands. 

In the CMD (left panel), $\log R_i$ exhibits a clear gradient with $i$-band magnitude: sources with fainter optical counterparts tend to have higher radio-to-optical ratios (see also Figure \ref{fig:i_logR_joint}).  
The high-$\log R_i$ population is not confined to the reddest locus in $g-i$;
instead, many such sources occupy intermediate colors and, in part, the blue cloud,
suggesting that optical emission in these radio-dominant systems can include a range of
host-galaxy and nuclear contributions.
We also find that many sources extend beyond the approximate UNIONS magnitude limit while spanning a broad range of $g-i$ colors, highlighting the ability of the deeper HSC imaging to probe the diverse optical colors of faint radio populations. 

The color--color diagram (middle panel) shows a prominent narrow ridge, plausibly associated with
the locus of typical galaxy SEDs as the 4000\,\AA\ break shifts through the
optical bands with redshift.
High-$\log R_i$ sources are more broadly distributed around this ridge, including a population extending toward relatively blue $g-i$ at intermediate $i-z$. 
This spread may reflect a combination of effects, including variations in stellar population age, dust attenuation, AGN contribution, and emission-line contamination, rather than a single dominant mechanism.

The $g-i$ color as a function of redshift (right panel) further clarifies this behavior.
At low redshift ($z \lesssim 0.5$), the color distribution follows a relatively tight sequence,
consistent with the red sequence of galaxies.
Toward higher redshift, the $g-i$ colors systematically increase up to $z \sim 0.8$--1,
reflecting the redshifting of the 4000\,\AA\ break through the optical bands.
At $z \gtrsim 1$, however, the color distribution becomes significantly broader,
indicating weaker constraints from the HSC $grizy$ photometry as the break shifts to longer wavelengths.
In this regime, high-$\log R_i$ sources span a wide range of colors,
suggesting a diversity of SEDs and a mixture of host-galaxy and AGN contributions.
These trends may reflect the presence of different accretion modes in the radio AGN population, commonly discussed in terms of high-excitation and low-excitation radio galaxies (HERGs and LERGs; e.g., \citealt{Buttiglione2010,BestHeckman2012,HardcastleCroston2020}).
HERGs are generally associated with radiatively efficient accretion and strong optical emission lines, whereas LERGs are typically linked to radiatively inefficient accretion and weak-line, jet-dominated systems.
The observed $g-i$ color trends are broadly consistent with the results of \citet{Ching2017}, who showed that HERGs and LERGs occupy different regions of optical color space, with HERGs exhibiting a broader and relatively bluer color distribution. 
Detailed HERG/LERG classifications using spectroscopic diagnostics will be explored in forthcoming spectroscopic studies.

\subsection{Radio Spectral Properties}

\begin{figure*}
  \centering
  \includegraphics[width=\textwidth]{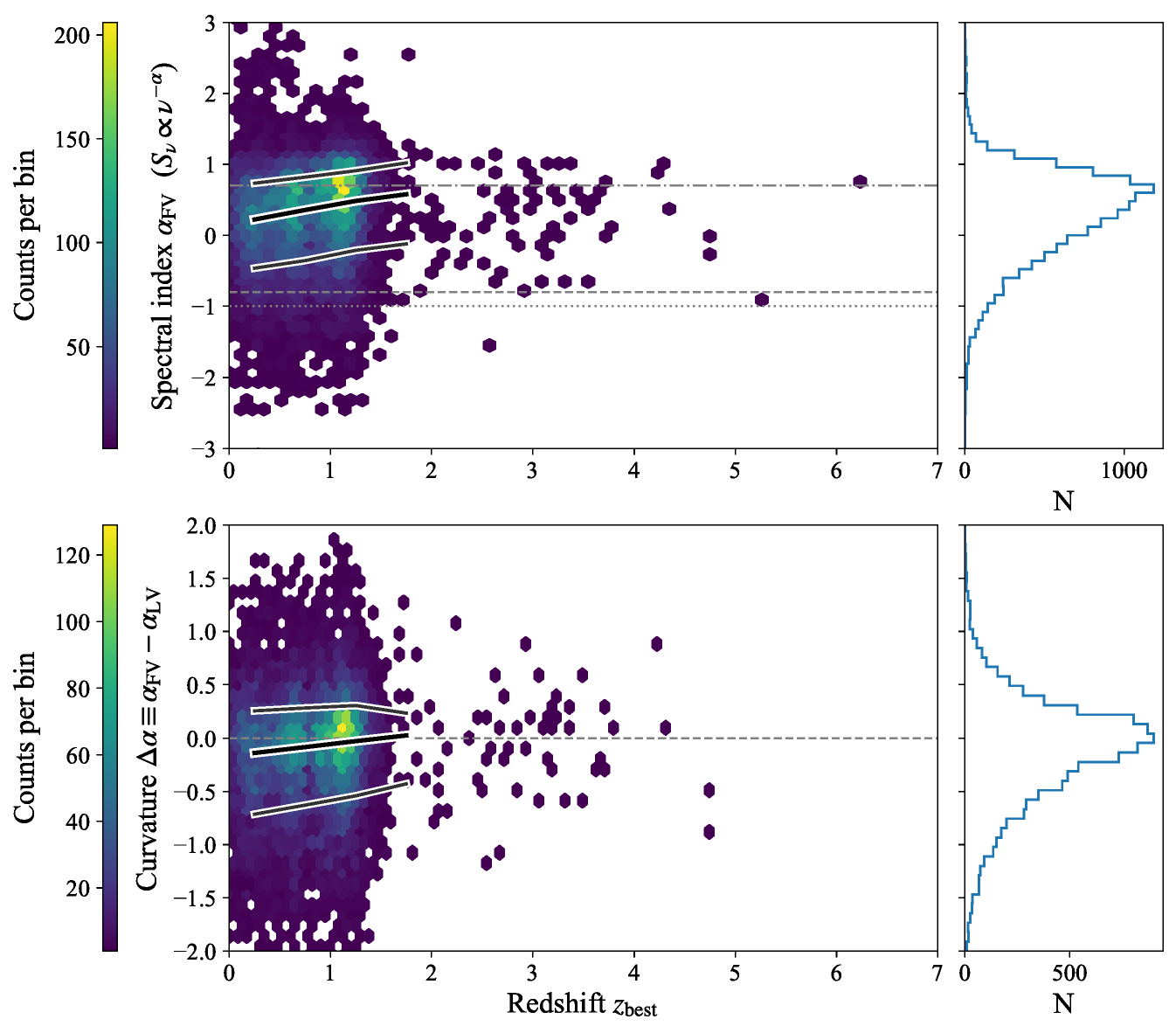}
\caption{
Spectral index and curvature as a function of photometric redshift for the Clean VLASS--HSC sample.
(Top) Hexbin map of $z_{\rm best}$ versus the two-point spectral index $\alpha_{\rm FV}$ measured between 1.4 and 3\,GHz (defined by $S_\nu\propto \nu^{-\alpha}$), with the marginal distribution shown on the right.
The horizontal lines indicate reference values, including the commonly adopted $\alpha=0.7$ (dot--dashed) and the ``steep-spectrum'' regime (dashed/dotted).
(Bottom) The corresponding redshift dependence of the spectral curvature, $\Delta\alpha\equiv \alpha_{\rm FV}-\alpha_{\rm LV}$; the dashed line marks $\Delta\alpha=0$. 
In both panels, the solid curves show the running median and the 16th/84th percentiles computed in redshift bins.
  \label{fig:z_alpha_curv}}
\end{figure*}

Figure~\ref{fig:z_alpha_curv} summarizes how the radio spectral index and curvature vary with redshift in our radio-selected sample.
Traditionally, high-redshift radio galaxies have often been efficiently pre-selected using the ultra-steep-spectrum (USS) criterion (e.g., $\alpha \gtrsim 1$), motivated by the empirical $z$--$\alpha$ correlation \citep[e.g.,][]{DeBreuck2000,Klamer2006SUMSS_NVSS_III}.
Our approach does not impose such a spectral-index cut, allowing us to assess how completely we recover sources across a wide range of $\alpha$.

The upper panel shows that, while a steep-spectrum population is certainly present, our sample spans a broad range of spectral indices, including comparatively flat spectra.
This confirms that the catalog is not restricted to USS-selected objects and therefore can capture radio AGN populations that would be under-represented in purely steep-spectrum searches (e.g., sources with a stronger compact/core contribution; \citealt{UrryPadovani1995}).
Interestingly, the running median of $\alpha_{\rm FV}$ shows a mild tendency to approach $\alpha\simeq 0.7$ at higher $z_{\rm phot}$.
A natural interpretation is that, as redshift increases in a flux-limited survey, the detected population becomes increasingly dominated by
powerful, lobe-dominated systems whose optically thin synchrotron spectra are well described by a near power-law with $\alpha\sim0.7$
(e.g., \citealt{Condon1992}), whereas at low $z$ the mixture includes a wider variety of spectral shapes.

The lower panel indicates that the spectral curvature, quantified by $\Delta\alpha$, is typically close to zero,
implying that many sources are approximately consistent with a single power law across 144\,MHz--3\,GHz.
We also find a weak tendency toward negative curvature at low redshift.
Such behavior can arise if the GHz-frequency spectrum is modestly flattened relative to the low-frequency baseline, for example due to
an increased fractional contribution from compact/core emission or due to non-simultaneous flux measurements between surveys,
both of which can reduce the apparent high-frequency slope \citep[e.g.,][]{UrryPadovani1995}.
In contrast, classical radiative ageing models predict curvature associated with synchrotron and inverse-Compton losses
\citep[e.g.,][]{Kardashev1962,JaffePerola1973};
the near-zero median curvature in our sample suggests that strong spectral breaks are not ubiquitous within the observed frequency range,
at least for the bulk of the sources.


\subsection{Morphological resolvability in HSC imaging} \label{sec:morph}

\begin{figure*}
  \centering
  \includegraphics[width=\textwidth]{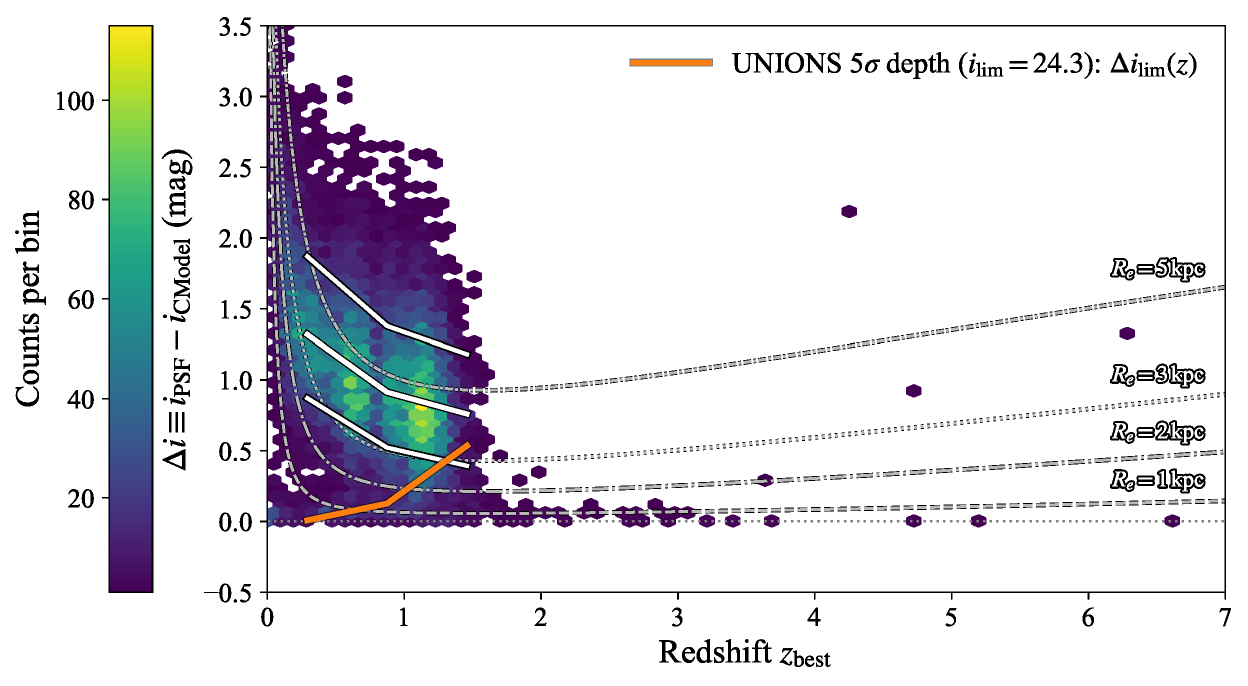}
\caption{
Morphology diagnostic for the extended subsample.
We plot the HSC $i$-band extendedness indicator
$\Delta i \equiv i_{\rm PSF}-i_{\rm CModel}$ as a function of the best-available redshift ($z_{\rm best}$). 
We apply an $i$-band signal-to-noise cut of ${\rm S/N}_i \ge 5$.  
Extended sources are then defined by a significance threshold $S_{\Delta i}\equiv \Delta i/\sigma_{\Delta i}\ge5$ and $\Delta i>0$,
with $\sigma_{\Delta i}=\sqrt{\sigma(i_{\rm PSF})^2+\sigma(i_{\rm CModel})^2}$;
only sources passing this extended criterion are shown.
Colors indicate the number of objects per hexagonal bin.
White solid curves show the 16th, 50th (median), and 84th percentiles of $\Delta i$ in $z_{\rm best}$ bins.
Gray curves indicate a simple theoretical expectation of $\Delta i$ for a Gaussian galaxy with
physical half-light radius $R_{\rm e}=1$--$5$\,kpc convolved with a Gaussian PSF of
${\rm FWHM}=0.7^{\prime\prime}$ \citep[a conservative value compared to the typical HSC $i$-band image quality, e.g.,][]{Aihara2018};
the conversion from physical size to angular size uses the angular-diameter distance in the Planck cosmology.
The orange solid curve provides a detectability guide for a shallower survey such as UNIONS:
assuming its nominal $i$-band depth $i_{\rm lim}\simeq24.3$ (5$\sigma$ point-source depth),
we estimate the magnitude-dependent uncertainty in $\Delta i$ via
${\rm S/N}(m)=5\times10^{-0.4(m-i_{\rm lim})}$ and $\sigma_{\Delta i}(m)\simeq\sqrt{2}\,(1.0857/{\rm S/N})$,
and plot the median $\Delta i_{\rm lim}(z)=5\,\sigma_{\Delta i}$ in redshift bins using the observed $i_{\rm CModel}$ distribution.
}
\label{fig:z_vs_deltai_extonly}
\end{figure*}

To illustrate how far HSC imaging can identify spatially extended counterparts at high redshift,
we use the $i$-band morphology diagnostic $\Delta i$ defined in \S\ref{subsec:catalog_deltai}.
We classify objects as extended when $S_{\Delta i}\ge 5$ and $\Delta i>0$, requiring ${\rm S/N}_i\ge 5$.


Figure~\ref{fig:z_vs_deltai_extonly} shows $z_{\rm best}$ versus $\Delta i$ for the extended subsample. 
A substantial population remains clearly extended around $z\sim1$, indicating that HSC imaging can identify resolved optical counterparts even at intermediate redshift. 
As a reference, we overlay simple theoretical curves for Gaussian sources with physical half-light radii of $R_{\rm e}=1$--$5$ kpc, convolved with a ${\rm FWHM}=0.7^{\prime\prime}$ PSF. 
The adopted PSF size is slightly more conservative than the typical $i$-band seeing of HSC ($\sim0.6^{\prime\prime}$; \citealt{Aihara2018}). 
Physical sizes are converted to angular sizes using the angular-diameter distance in the Planck cosmology. 
Although real galaxies are not perfect Gaussians, these curves provide an intuitive guide to how $\Delta i$ depends on source size and redshift.

We also indicate a detectability guide for a shallower survey such as UNIONS, whose nominal depth is
$i_{\rm lim}\simeq24.3$ (5$\sigma$ point source in a 2$^{\prime\prime}$ aperture). 
To account for the fact that the uncertainty in $\Delta i$ depends on source brightness, we model the
$i$-band signal-to-noise at magnitude $m$ as ${\rm S/N}(m)=5\times10^{-0.4(m-i_{\rm lim})}$, which implies a representative magnitude uncertainty $\sigma_m(m)\simeq1.0857/{\rm S/N}(m)$.
Assuming independent errors for $i_{\rm PSF}$ and $i_{\rm CModel}$, we approximate
$\sigma_{\Delta i}(m)\simeq \sqrt{2}\,\sigma_m(m)$ and thus
$\Delta i_{\rm lim}(m)\simeq S_{\Delta i}\,\sigma_{\Delta i}(m)$ for a chosen significance threshold $S_{\Delta i}$. 
The curve shown in Figure~\ref{fig:z_vs_deltai_extonly} is obtained by evaluating $\Delta i_{\rm lim}(m)$ using each object's observed
$i_{\rm CModel}$ magnitude and plotting the median $\Delta i_{\rm lim}$ in bins of $z_{\rm best}$ (with $S_{\Delta i}=5$),
illustrating how modest extensions detectable in HSC can be missed at shallower depth.
Our morphology selection preferentially highlights host-dominated AGN candidates, which are expected to include a substantial fraction of obscured \citep[type-2 like; ][]{UrryPadovani1995} systems, although morphology alone does not uniquely determine the AGN type. 

At high-$z$, spectroscopic follow-up is often biased toward quasar-like targets, for which secure redshifts can be obtained efficiently from prominent broad emission lines. 
Consistent with this selection, the $z\gtrsim2$ objects in our radio-galaxy sample that have spectroscopic redshifts are predominantly radio-loud quasars, and thus appear optically compact (small $\Delta i$) in the $\Delta i$--$z$ plane.

\subsection{Comparison with the first WERGS (WERGS I) catalog of Yamashita et al.\ (2018)} \label{subsec:comp_yamashita2018}

Our catalog builds on the original WERGS I effort presented by \citet{Yamashita2018WERGS1}, which demonstrated the power of deep HSC imaging for identifying optical counterparts to radio sources.
In WERGS~I, a $1''$ positional cross-match between FIRST and early HSC-SSP data ($i\lesssim26$) over 154~deg$^{2}$ yielded $>3600$ optical counterparts, corresponding to $>50\%$ of the FIRST sources in the search footprint \citep{Yamashita2018WERGS1}.
They further showed that the HSC depth substantially increases the counterpart recovery relative to SDSS-depth ($i\sim21$) identifications  and that the matched radio-galaxy population is dominated at $z\lesssim1.5$ based on $grizy$ photometric redshifts \citep{Yamashita2018WERGS1}.

The present work extends this WERGS-style approach in three key ways.
First, we construct a homogeneous primary catalog over the final-year HSC-SSP Wide footprint ($\approx1200$~deg$^{2}$) using the latest internal processing (DR S23B), providing a substantially larger optical foundation than the 154~deg$^{2}$ HSC-SSP DR S16A footprint. 
As a result, our primary Clean VLASS--HSC catalog contains 22{,}773 sources.
This corresponds to an increase by a factor of $\sim6.3$ in the number of optically identified radio sources relative to the WERGS~I counterpart sample. 
At $z_{\rm best}>2$, our Clean VLASS--HSC catalog includes 81 sources, an increase from the $\sim$20 objects reported in WERGS~I \citep[][]{Yamashita2018WERGS1}, highlighting our improved recovery of high-redshift radio sources enabled by the wider HSC footprint.
In addition, the number of high-$z$ candidates will increase by not relying solely on the photo-$z$ methods, but instead by applying dropout selections in a forthcoming paper (WERGS XIII; Kong et al. submitted). Using this approach, we identify $\sim400$ high-$z$ candidates at $z>4$, thanks to the expanded wide area footprint of $\approx1200$~deg$^2$. 

Second, we adopt the VLASS Epoch~2 3\,GHz component catalog as the parent radio sample. 
Its $\sim$2$^{\prime\prime}$ resolution (compared to the $\sim$6$^{\prime\prime}$ resolution of FIRST) enables more precise positional associations, reducing chance matches and permitting smaller matching radii. 
In fact, Figure~\ref{fig:sep_hist_3panel} and Table~\ref{tab:sep_stats} show that the typical radio--optical separation is substantially smaller for VLASS than for FIRST: the median nearest-neighbor separation is $0\farcs199$ for VLASS--HSC, compared to $0\farcs34$ for FIRST--HSC. 
This improvement reflects the higher-precision radio positions in VLASS, enabling more reliable optical associations at a given matching radius.

Third, we provide consistent cross-match information to FIRST (1.4 GHz) and LoTSS (144 MHz), enabling straightforward radio multi-frequency analysis. 
In particular, the broad frequency leverage (144 MHz--3 GHz) allows more robust constraints on radio spectral slopes and curvature, which in turn improves K-corrections and the derivation of rest-frame radio luminosities. 
As shown in Figure~\ref{fig:z_alpha_curv}, the mean spectral index is close to the canonical value ($\langle\alpha\rangle \approx -0.7$), while the distribution displays large scatter, extending from $\alpha_{\rm FV}\sim-2$ to $\sim+2$. 
The extreme tails may reflect a combination of measurement uncertainties, resolution/matching effects, variability, and genuine spectral curvature (e.g., GPS/CSS). 
Nevertheless, the broad dispersion highlights that assuming a single $\alpha$ can bias K-corrections for individual objects, reinforcing the value of our multi-frequency cross-matches.


\section{Conclusion}
\label{sec:summary}

We present a wide-area optical identification catalog for radio sources based on VLASS Epoch~2 SE components and the final-year internal processing of the Subaru HSC-SSP Wide layer (DR S23B).
Our primary product is the Clean VLASS--HSC catalog containing 22,773 sources with robust HSC counterparts, $grizy$ photometry. 
We additionally provide ancillary nearest-neighbor associations to FIRST (1.4\,GHz) and LoTSS (144\,MHz), enabling straightforward multi-frequency analyses without imposing spectral-index pre-selection. 
Key results of this work are as follows:

\begin{enumerate}
\item We define the primary VLASS--HSC catalog via a $1\farcs0$ positional match and conservative optical quality cuts (S/N$>5$ in at least one band), yielding 22,773 sources over the HSC-SSP Wide footprint. 
\item Ancillary matches to FIRST and LoTSS using $2\farcs5$ radii yield 18,444 FIRST-matched sources and 16,167 LoTSS-matched sources; 14,206 sources have counterparts in both surveys, providing a well-defined multi-frequency subset for spectral-index and radio-color studies. 
\item We extend the WERGS-style cross-matching in three ways: (i) we build a homogeneous primary catalog over the final-year HSC-SSP Wide footprint ($1200~\mathrm{deg}^{2}$; internal DR~S23B), yielding a Clean VLASS--HSC sample of 22{,}773 sources ($\sim 6.3$ times larger than WERGS~I) and 81 sources at $z>2$; (ii) we adopt the VLASS Epoch~2 3\,GHz component catalog, whose $\sim$2$^{\prime\prime}$ resolution enables more precise optical associations (median VLASS--HSC separation $0\farcs199$ versus $0\farcs34$ for FIRST--HSC); and (iii) we provide uniform cross-match information to FIRST (1.4\,GHz) and LoTSS (144\,MHz), enabling multi-frequency spectral-index/curvature measurements and improved radio K-corrections and rest-frame luminosities.
\item Compared to UNIONS-based optical identifications of VLASS sources \citep{Zhong2025UNVEIL1},
our catalog covers a smaller area but reaches substantially deeper optical photometry ($i\sim26$ for HSC-SSP Wide versus $i\sim24.3$ for UNIONS, i.e.\ deeper by $\sim1.5$--1.7~mag),
thereby improving sensitivity to optically faint hosts.
This depth gain enhances the completeness for extremely radio-dominant systems at fixed radio flux density and expands discovery space for rare, high radio-loudness populations.
In addition, the deeper HSC imaging enables meaningful morphological discrimination, and we find that a non-negligible fraction of radio-selected hosts are spatially resolved even at $z\gtrsim1$, which would be difficult to achieve at the shallower UNIONS depth. 
\end{enumerate} 

Future spectroscopic follow-up of these radio-selected galaxies will be enabled by the Subaru Prime Focus Spectrograph Subaru Strategic Program (PFS-SSP; \citealt{Takada2014PFS}),
which will deliver wide-area, multiplexed optical/near-infrared spectroscopy and provide large statistical samples for galaxy/AGN evolution studies (see also \citealt{Greene2022PFS_GE}).

\acknowledgments
We thank the anonymous referee and the data editor for their constructive comments, which helped improve the manuscript and associated data products. 
This work was supported by the Japan Society for the Promotion of Science (JSPS) KAKENHI (JP22K14075, JP24K00684, JP25H00663 (H.U.); JP25K01043 (K.I.); JP23K20035 and JP24H00004 (K.K.)); JP23H01215 and JP25H00671 (T.N.) 
K.I. also acknowledges support from the JST FOREST Program, Grant Number JPMJFR2466 and the Inamori Research Grants, which helped make this research possible.

The Hyper Suprime-Cam (HSC) collaboration includes the astronomical communities of Japan and Taiwan, and Princeton University.  The HSC instrumentation and software were developed by the National Astronomical Observatory of Japan (NAOJ), the Kavli Institute for the Physics and Mathematics of the Universe (Kavli IPMU), the University of Tokyo, the High Energy Accelerator Research Organization (KEK), the Academia Sinica Institute for Astronomy and Astrophysics in Taiwan (ASIAA), and Princeton University.  Funding was contributed by the FIRST program from the Japanese Cabinet Office, the Ministry of Education, Culture, Sports, Science and Technology (MEXT), the Japan Society for the Promotion of Science (JSPS), Japan Science and Technology Agency  (JST), the Toray Science  Foundation, NAOJ, Kavli IPMU, KEK, ASIAA, and Princeton University.
This paper is based [in part] on data collected at the Subaru Telescope and retrieved from the HSC data archive system, which is operated by Subaru Telescope and Astronomy Data Center (ADC) at NAOJ. Data analysis was in part carried out with the cooperation of Center for Computational Astrophysics (CfCA) at NAOJ.  We are honored and grateful for the opportunity of observing the Universe from Maunakea, which has the cultural, historical and natural significance in Hawaii.

This paper makes use of software developed for Vera C. Rubin Observatory. We thank the Rubin Observatory for making their code available as free software at http://pipelines.lsst.io/. 

The Pan-STARRS1 Surveys (PS1) and the PS1 public science archive have been made possible through contributions by the Institute for Astronomy, the University of Hawaii, the Pan-STARRS Project Office, the Max Planck Society and its participating institutes, the Max Planck Institute for Astronomy, Heidelberg, and the Max Planck Institute for Extraterrestrial Physics, Garching, The Johns Hopkins University, Durham University, the University of Edinburgh, the Queen’s University Belfast, the Harvard-Smithsonian Center for Astrophysics, the Las Cumbres Observatory Global Telescope Network Incorporated, the National Central University of Taiwan, the Space Telescope Science Institute, the National Aeronautics and Space Administration under grant No. NNX08AR22G issued through the Planetary Science Division of the NASA Science Mission Directorate, the National Science Foundation grant No. AST-1238877, the University of Maryland, Eotvos Lorand University (ELTE), the Los Alamos National Laboratory, and the Gordon and Betty Moore Foundation.

\bibliographystyle{aasjournal} 
\bibliography{reference}

\appendix
\section{Justification of the $1\farcs0$ VLASS--HSC matching radius}
\label{app:match_radius}

In this Appendix, we estimate the expected number of chance VLASS--HSC associations as a function of matching radius, and quantify the contamination fraction in our positional cross-match.
Our approach is intentionally empirical: we measure the surface density of HSC sources that pass our quality cuts in a representative subset of the survey footprint, and combine it with the observed number of VLASS--HSC matches to infer the fraction of spurious associations.

\subsection{Estimating the HSC source surface density}
\label{app:nhsc_sigma}

We estimate the surface density of HSC objects, $\Sigma_{\rm HSC}$ (arcsec$^{-2}$), after applying the same HSC-side quality cuts as used for our optical counterparts (Table \ref{tab:hsc_clean_flags}).
To sample the footprint efficiently, we first draw 20 tracts at random from the Wide layer. 
For each tract, we query the HSC forced-photometry table (\nolinkurl{s23b_wide.forced}) 
, enforcing the quality cuts and additionally requiring a $\mathrm{S/N}>5$ detection in at least one band ($g,r,i,z,$ or $y$), and then count the resulting objects.
We denote the total number of such HSC objects in the 20 tracts as $N_{\rm HSC}$.

To convert this count into a surface density, we estimate the effective unmasked area, $A_{\rm eff}$, over the same set of tracts.
We query the HSC random catalog (\nolinkurl{s23b_wide.random}) with the identical mask and pixel-flag criteria, and count the number of surviving random points, $N_{\rm rand}$.
Because the HSC random catalog is constructed with a known areal density (100 random points per arcmin$^2$), the effective area is
\begin{equation}
A_{\rm eff} = \frac{N_{\rm rand}}{100} \ {\rm arcmin}^2.  
\end{equation}
We then compute
\begin{equation}
\Sigma_{\rm HSC} = \frac{N_{\rm HSC}}{A_{\rm eff}} \ \ ({\rm arcsec}^{-2}) .
\end{equation}

\subsection{Chance-coincidence model and contamination fraction}
\label{app:chance_model}

Given $\Sigma_{\rm HSC}$, the expected number of unrelated HSC objects within a radius $r$ of a VLASS position is
\begin{equation}
\lambda(r) = \Sigma_{\rm HSC}\,\pi r^2 .
\end{equation}
Assuming a Poisson distribution for background sources, the probability that a VLASS source has at least one chance HSC counterpart within $r$ is
\begin{equation}
P(\ge 1; r) = 1 - e^{-\lambda(r)} .
\end{equation}
For the VLASS sample restricted to the effective overlap area, with size $N_{\rm sample}$, the expected number of chance matches is
\begin{equation}
N_{\rm chance}(r) = N_{\rm sample}\,P(\ge 1; r) .
\end{equation}

\begin{figure}
  \centering
  \includegraphics[width=0.5\linewidth]{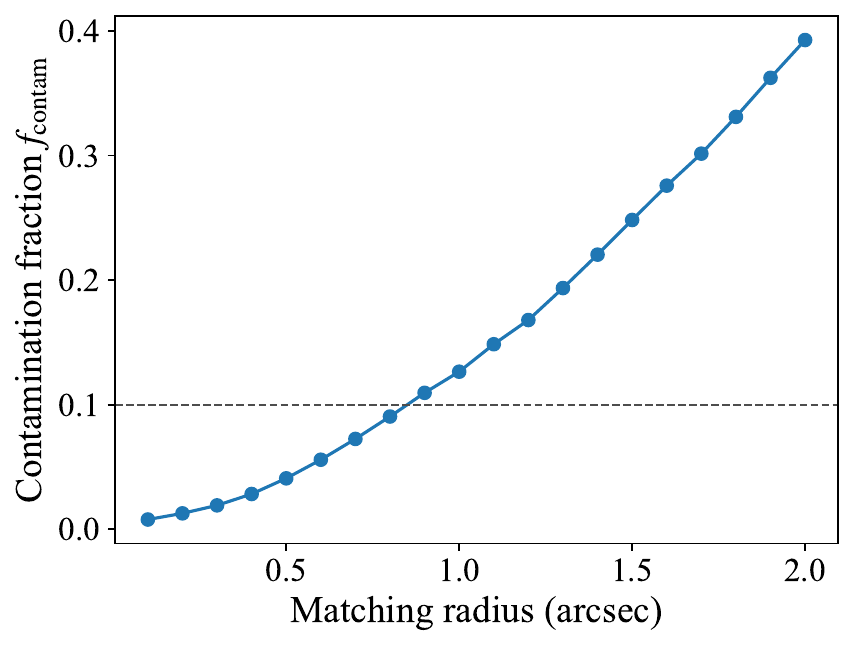}
  \caption{Estimated contamination fraction, $f_{\rm contam}$, as a function of the VLASS--HSC matching radius.}
  \label{fig:contam}
\end{figure}

We measure the actual number of matches, $N_{\rm matched}(r)$, by performing a nearest-neighbor match between VLASS sources (restricted to the same effective area) and the HSC catalog, and counting matches within radii from $0\farcs1$ to $2\farcs0$ in $0\farcs1$ steps.
Finally, we estimate the contamination fraction as
\begin{equation}
f_{\rm contam}(r) = \frac{N_{\rm chance}(r)}{N_{\rm matched}(r)} .
\end{equation}
We find that $f_{\rm contam}(1\farcs0) \lesssim 0.1$, motivating our adoption of a $1\farcs0$ matching radius as shown in Figure \ref{fig:contam}.

\section{Number of VLASS sources within the HSC--SSP Wide footprint}

To estimate how many sources from the Clean VLASS catalog fall within the HSC--SSP Wide footprint, we perform a positional search around each
VLASS source using the HSC catalog.

First, we search for HSC objects within a radius of $60''$ around each VLASS source without applying the bright-star masks or other HSC quality cuts.
This yields $62,867$ VLASS sources within the nominal HSC--SSP Wide imaging footprint.

Next, we repeat the same search after applying the HSC-side selection used to define the clean optical sample, except for the \texttt{sdsscentroid} requirement in Table~\ref{tab:hsc_clean_flags}.
Specifically, we require primary detections, excluded edge/interpolated, saturated, cosmic-ray, and bad-pixel affected measurements, applied the bright-star masks, and imposed the input-count cuts.
A VLASS source is then considered to lie within the effective HSC footprint if at least one HSC object satisfying these criteria is found
within $60''$.

Applying these masking and quality criteria reduces the number of VLASS sources within the effective HSC footprint to $44,066$.
For comparison, the full Clean VLASS catalog contains $1,380,707$ sources, implying that approximately $3.2\%$ of the Clean VLASS sources
fall within the effective HSC--SSP Wide survey area.

\end{document}